\documentclass[11pt]{article}
\usepackage[margin=1.2in]{geometry}
\usepackage[utf8]{inputenc}
\usepackage[T1]{fontenc}
\usepackage{authblk}
\usepackage{enumitem}

\usepackage{amsmath} 
\usepackage{amsthm}
\usepackage{amsfonts}
\usepackage{mathtools}

\usepackage{graphicx}
\usepackage{url}
\usepackage{subcaption}
\usepackage[hyperfootnotes=false,colorlinks=true]{hyperref}

\usepackage{tikz}
\usetikzlibrary{calc}
\usetikzlibrary{backgrounds}
\usetikzlibrary{intersections}
\usetikzlibrary{decorations.pathreplacing}
\usetikzlibrary{decorations.pathmorphing}
\usetikzlibrary{decorations.markings}
\usetikzlibrary{arrows}
\usetikzlibrary{arrows.meta}

\theoremstyle{plain}
\newtheorem{theorem}{Theorem}

\theoremstyle{definition}

\newtheorem{example}{Example}

\DeclareRobustCommand\tikzCross{\tikz {\draw(.15, .15) -- (-.15, -.15); \draw(-.15, .15) -- (.15, -.15);} }

\title{Catacondensed Chemical Hexagonal Complexes: \\
A Natural Generalisation of Benzenoids
}

\author[1]{Cate~S.~Anst{\"o}ter}
\author[2,3,4]{Nino Ba{\v s}i{\'c}}
\author[5]{Patrick~W.~Fowler}
\author[2,3,4,6]{Toma{\v z} Pisanski}

\affil[1]{Department of Chemistry, Temple University, Philadelphia, 19122, USA}
\affil[2]{FAMNIT, University of Primorska, Koper, Slovenia}
\affil[3]{IAM, University of Primorska, Koper, Slovenia}
\affil[4]{Institute of Mathematics, Physics and Mechanics, Ljubljana, Slovenia}
\affil[5]{Department of Chemistry, University of Sheffield, Sheffield S3 7HF, UK}
\affil[6]{Faculty of Mathematics and Physics, University of Ljubljana, Ljubljana, Slovenia}

\date{March 31, 2021 (revised on May 27, 2021)}

\begin{document}

\maketitle

\vspace{-\baselineskip}
\centerline{\bfseries Dedicated to Milan Randić on the occasion of his 90th birthday.} 

\begin{abstract}
Catacondensed benzenoids (those benzenoids having
no carbon atom belonging to three hexagonal rings)
form the simplest class of polycyclic aromatic
hydrocarbons (PAH).
They have a long history of study and are of
wide chemical importance.
In this paper,
mathematical possibilities for natural extension of the notion
of a catacondensed benzenoid are discussed, leading under plausible chemically and physically
motivated restrictions to the notion of a
{\it catacondensed chemical hexagonal complex} (CCHC).
A general \emph{polygonal complex} is a topological structure composed of polygons that are glued together along certain edges.
A polygonal complex is \emph{flat} if none of its edges belong to more than two polygons. A connected flat polygonal complex determines
an orientable or nonorientable surface, possibly with boundary.
A CCHC is then a connected flat polygonal complex all of whose polygons are hexagons and each of whose vertices belongs to
at most two hexagonal faces.
We prove that all CCHC are Kekulean and give formulas for
counting the perfect matchings in a series of examples based on expansion of cubic graphs
in which the edges are replaced by linear polyacenes of equal length.
As a preliminary assessment of the likely stability of molecules with CCHC structure,
all-electron quantum chemical calculations are applied to molecular structures based
on several CCHC, using either linear or kinked unbranched catafused polyacenes as the expansion motif.
The systems examined all have closed shells according to H\"uckel theory and all
correspond to minima on the potential surface, thus passing the most basic test for plausibility as a chemical species.
Preliminary indications are that relative energies of isomers are affected by the choice of the catafusene motif,
with a preference shown for kinked over linear polyacenes, and for attachment by
angular connection at the branching hexagons derived from the vertices of the underlying cubic
structure. Avoidance of steric crowding of H atoms appears to be a significant factor in these preferences.

\vspace{\baselineskip}
\noindent
\textbf{Keywords:} Benzenoid, polygonal complex,
(catacondensed) chemical hexagonal complex, Kekul\'{e} structure.

\vspace{\baselineskip}
\noindent
\textbf{Math.\ Subj.\ Class.\ (2020):} 
05C92 % Chemical graph theory
\end{abstract}

\section{Introduction} 

The familiar classes of conjugated unsaturated hydrocarbon molecules, such as
benzenoids, coronoids, helicenes and more general fusenes, may all be regarded 
in a mathematical sense as sets of graphs equipped
with additional properties. 
In the simplest case, the hexagonal rings of such molecules
may be considered as faces of a map on the plane.
In this note we extend this notion by retaining the local
properties of benzenoids but relaxing global planarity. 
Since the first isolation of benzene almost 200 years ago, 
benzenoids and their derivatives have had a significant, if not always 
benign, presence in the mainstream of organic chemistry and its applications.
Mathematical study of benzenoids also has a long history, with central
ideas \cite{Crocker1922,Armit1925} contributed by pioneering experimental
chemists such as 
Kekul{\'e} \cite{Kekule1865}, 
Fries \cite{Fries1927}
and Clar \cite{clar1964,clar1964a,Clar1972}
feeding into an enormous primary
literature codified in influential textbooks \cite{cyvin_1988,
cyvin1991,cyvin1994,gutman1989,trinajstic1992}.
The present paper is dedicated to another major figure,
Milan Randi\'{c}, whose ideas on the use of conjugated circuits \cite{Randic1976Ccar} 
for the description of resonance energy have been
influencing thinking in this area for
nearly half a century \cite{RaNiTr1987,ElRa1987,Ra1991,Ra1993,PiRa2000,GuRa2001,VuRaBa2004,BaRa2005,BaRa2007,BaRaVu2008,RaNoPl2012}. 
Most recently, his simple but insightful
picture of ring-current aromaticity of benzenoids has 
revived interest in ways of modelling and, especially, of interpreting molecular currents \cite{Gomes1979,Randic2010Gtat,
Rand2011,Rand2012A,Rand2012B,Rand2012C,Cies2009,Mandado2009,PWF2011,PWF2016,PWF2020}.
 
The simplest structures, with many applications in chemistry, are \emph{benzenoids}. 
Graphene
may be viewed as an infinite benzenoid. In this paper we are interested only in finite structures.
There are several ways to describe a finite benzenoid: boundary-edges codes \cite{guo2002,BasFowPis2017,deza1990}, inner duals \cite{Nikolic1990,Mallion1983}, 
flag-graphs \cite{kovic_2014,balaban2012}, or through the coordinates of the hexagons
in the infinite hexagonal tesselation of the plane \cite{Bas2017}. Benzenoids having no inner vertices
(i.e.\  no vertices common to three hexagons) 
are called \emph{catacondensed}, whilst those having inner vertices are called \emph{pericondensed}.
Among catacondensed benzenoids we distinguish \emph{branched} and 
\emph{unbranched} benzenoids. The simplest unbranched benzenoids are \emph{linear benzenoids} or \emph{linear polyacenes}. 

If the structures are allowed to spiral we call them\emph{ helicenes}. These are still planar,
in the graph theoretical sense,
 and simply
connected but no longer fit onto a hexagonal grid without overlap. The term \emph{fusenes}
covers both benzenoids and helicenes. Note that the boundary does not determine uniquely a general fusene; 
see for instance work by Brinkmann \cite{Brinkmann2002,Brinkmann2007}.

Benzenoids with \emph{holes} (i.e.\ those that are not simply connected) are \emph{coronoids}. Again, those that have
no internal vertices are catacondensed coronoids (or perhaps more simply, \emph{catacoronoids} to correspond to \emph{catabenzenoids}).
 Benzenoids and coronoids have both been considered as maps on a surface with boundary \cite{kovic_2014,balaban2012,Basic2016}.
Fusenes can be further generalised to allow for structures that are not necessarily simply connected.
In the literature, various generalizations to  surfaces of higher genus have been made. For instance, \emph{torusenes} 
(also called  toroidal polyhexes or torenes) have been considered \cite{KiMaPo1993,MaPi2000,KiPi2007}. Since we may tile the Klein bottle by hexagons \cite{Deza2000}, we
may also speak of \emph{kleinbottlenes}. 
There is a whole menagerie of proposed finite and infinite theoretical carbon nanostructures,
such as M\"{o}biusenes, tubulenes, hexagonal systems, hexagonal animals, toroidal benzenoids, 
Schwarzites, Haeckelites,
etc.\ \cite{Gu1984,Sa1984,
SaHaZh1996,TrZi2015,TrZi2015a,Tr2017,Terrones1993,Terrones2000}.
The theory of maps \cite{Gross1987,PiPo2004} offers a toolbox for a general treatment of these
diverse structures.

Note that each map on a surface determines a graph, called the \emph{skeleton} of the map, that is obtained by discarding the faces of the map and retaining the 
vertices and edges. Whilst the skeleton is uniquely determined 
by the map, the converse is not true. A given graph may be a skeleton of several non-isomorphic maps. 
This fact has long been known to geometers:
% well-known to geometers of the nineteenth century and probably much earlier: 
 it was already Johannes Kepler who
presented non-convex regular polyhedra \cite{Kepler1997}. For instance, the great dodecahedron has the same skeleton as the icosahedron. 
Another example is the skeleton of 
the tetrahedron, which is the complete graph $K_4$. The graph $K_4$ is also the skeleton of the hemihexahedron (also called hemicube), a map with 
three quadrilateral faces in the projective plane \cite{Coxeter1973,McMullen2002}.
In mathematical chemistry this problem is relevant when counting the number of distinct toroidal polyhexes. One has 
to choose whether to count graphs or maps. Pisanski and Randić \cite{PiRa2000} give the example of the cube graph ($Q_3$), which has two non-equivalent hexagonal embeddings in the torus;
see also Figure~\ref{fig:threeembeddings} below.

In the next section, we present a flexible language for describing benzenoids and their many generalisations.

\section{Polygonal complex}

\subsection{Scheme}

Following Ringel \cite{Ri1974}, one can describe a cellular embedding of a graph in a closed surface by a \emph{scheme}.
Here we generalise Ringel's approach in two directions. If we do not insist that each symbol appears exactly twice, we
may use such schemes to describe the combinatorial structure of more general polygonal complexes in the sense of
Schulte \emph{et al.}~\cite{PeSc2010,PeSc2013,PeSc2014,ScWe2017}. 
On the other hand, if we allow symbols with a single appearance, we may describe chemical structures, such as benzenoids
as graphs embedded in a surface with a boundary.

Assume we are given a finite alphabet $A$. To each symbol $a \in A$ assign two literals $a^+, a^-$.  
We say that $a^+$ is inverse of $a^-$ and that $a^-$ is inverse of $a^+$.
Hence, if alphabet 
$A$ has $n$ symbols, there are $2n$ literals. 
When there is no ambiguity, we will write $a$ for $a^+$. 
A \emph{word} over literals denotes an oriented polygon. A sequence of words, also called a \emph{scheme}, denotes a \emph{polygonal complex}, i.e.\ collection 
of polygons, some glued along their edges. A double appearance of a symbol represents the gluing. If the symbols appear 
in the same literal, the gluing is \emph{parallel}; otherwise it is \emph{antiparallel}. 
This terminology is used in the description of polyhedral self-assembly in synthetic biology \cite{FiPiRu2014,Koetal2016}.
Usually, we present a scheme in a tabular form, where each row corresponds to a word.

Ringel \cite{Ri1974} defines some operations on schemes that induce an equivalence relation 
such that two equivalent schemes define the same polygonal complex. 
Two schemes are \emph{equivalent} if one can be obtained from the other by a sequence of transformations of the following types:
\begin{enumerate}[label=(T\arabic*)]
\item \label{T:1} Permute the rows of a scheme (since we may always reorder the list of polygons);
\item \label{T:2} Make a cyclic permutation of a row (since we may always start following the edges of a polygon from any of its vertices);
\item \label{T:3} Replace any symbol by an unused symbol while keeping the exponents (since we may always relabel the edges of the polygonal complex); 
\item \label{T:4} Pick a symbol and replace each occurrence of a literal by its inverse (since we may always reverse the direction of any edge);
\item \label{T:5} Reverse the row and simultaneously replace each literal by its inverse (since we may always reverse the orientation of any polygon).
\end{enumerate}

A scheme may satisfy some additional properties. For example:
\begin{enumerate}[label=(S\arabic*)]
\item \label{prop:connected}
A scheme is \emph{connected} if it cannot be divided into two disjoint sub-schemes that have no
symbol in common;
\item \label{prop:flat}
A scheme is \emph{flat} if each symbol appears at most twice in the scheme;
\item \label{prop:closed}
A scheme is \emph{closed} if each symbol appears at least twice in the scheme; 
\item \label{prop:linear}
A scheme is \emph{linear} if each word that contains exactly two symbols that appear multiple times in the scheme has them in antipodal positions;
\item \label{prop:chemical}
A scheme is \emph{chemical} if, whenever $ab$ appears in the scheme such that $a$ and $b$ both have multiple appearance, then there exists a literal $c$ (different 
from $a$ and $b$) such that $b^-c$ (or, alternatively, $ c^-b $) and $c^-a $ (or, alternatively, $ a^-c $) 
appear in the same scheme; 
\item \label{prop:catacondensed}
A scheme is \emph{catacondensed} if, whenever $ ab $ appears in the scheme, then at least one of the symbols $a$ and $b$ appears only once; 
\item \label{prop:unbranched}
A scheme is \emph{unbranched} if every word of the scheme contains at most two symbols that appear more than once in the scheme and 
if there are two in a given word, they are non-adjacent. A catacondensed scheme is called \emph{branched} whenever it is not unbranched;
\item \label{prop:hexagonal}
A scheme is \emph{hexagonal} if each word contains six literals;
\item \label{prop:oriented}
A scheme is \emph{oriented} if no literal appears in it twice (i.e.\ no symbol appears twice with the same exponent). It is \emph{orientable} if it is equivalent to an oriented scheme.
A scheme that is not orientable is \emph{nonorientable}.
\end{enumerate}
Note that \ref{prop:linear} implies \ref{prop:unbranched} and \ref{prop:unbranched} implies \ref{prop:catacondensed}. Also, every orientable scheme is flat.
All properties \ref{prop:connected} -- \ref{prop:hexagonal} are preserved under the aforementioned transformations \ref{T:1} -- \ref{T:5} and hence also apply to polygonal complexes.
The property \ref{prop:chemical} is equivalent to requiring that the skeleton graph is a chemical graph (i.e.\ has maximum degree less than or equal to $3$).
The property `oriented' \ref{prop:oriented} is not preserved under \ref{T:5}, though it is still preserved under \ref{T:1} -- \ref{T:4}. However,
properties `orientable' and `nonorientable' are preserved under \ref{T:1} -- \ref{T:5}.
We know that a connected flat scheme represents a compact surface with a boundary. The boundary is determined by 
symbols that appear only once in the scheme. 
If a connected flat scheme is also closed then the surface itself is closed, i.e.\ it has no boundary.
With these definitions, a fullerene is a case of a closed chemical complex that is not hexagonal,
since it has $12$ pentagonal faces \cite{atlas2007}.

\begin{example}
\label{exam:1}
A typical example of a polygonal complex in the sense of Schulte \emph{et al.}~\cite{PeSc2010,PeSc2013,PeSc2014,ScWe2017}
is a $2$-dimensional skeleton of the 
tesseract (the $4$-dimensional cube). This skeleton is composed of $16$ vertices, $32$ edges and $24$ quadrilateral faces. 
\begin{figure}[!ht]
\centering
\includegraphics[scale=0.4]{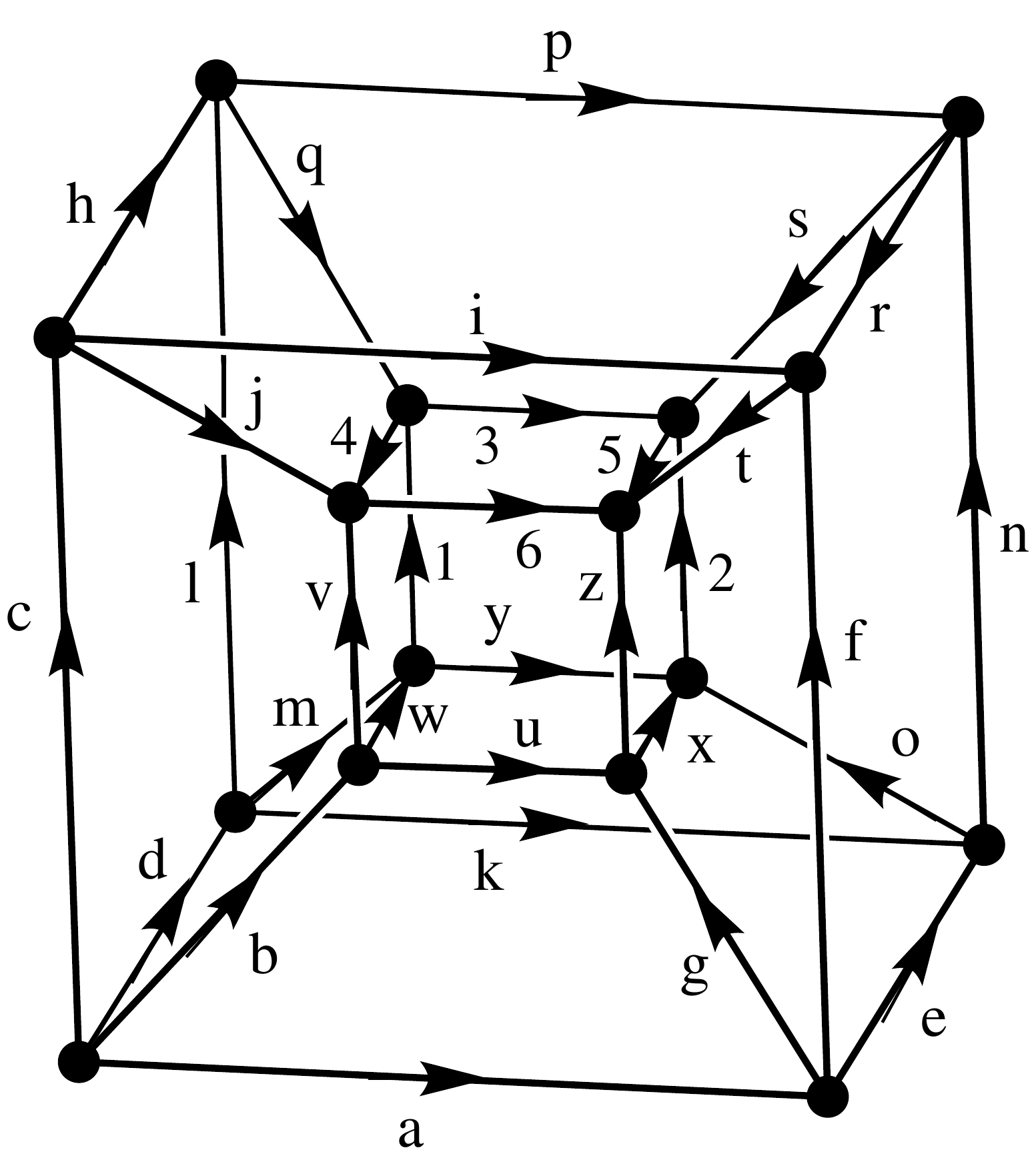}
\caption{The tesseract showing edge labelling as in the scheme presented in Example \ref{exam:1}.}
\label{fig:tesseract}
\end{figure}
The eight facets of the tesseract (which are all cubes) are discarded. A scheme describing the skeleton is given here (split into two columns for convenience):
\begin{align}
& \begin{matrix*}[l]
a & f & i^- & c^- \\
a & e & k^- & d^- \\
a & g & u^- & b^- \\
b & v & j^- & c^- \\
b & w & m^- & d^- \\
c & h & l^- & d^- \\
e & n & r & f^- \\
e & o & x^- & g^- \\
f & t & z^- & g^- \\
h & p & r & i^- \\
h & q & 4 & j^- \\
i & t & 6^- & j^- 
\end{matrix*} & 
& \begin{matrix*}[l]
k & n & p^- & l^- \\
k & o & y^- & m^- \\
l & q & 1^- & m^- \\
n & s & 2^- & o^- \\
p & s & 3^- & q^- \\
r & t & 5^- & s^- \\
u & x & y^- & w^- \\
u & z & 6^- & v^- \\
v & 4^- & 1^- & w^- \\
z & 5^- & 2^- & x^- \\
3 & 5 & 6^- & 4^- \\
1 & 3 & 2^- & y^- 
\end{matrix*}
\end{align}

As we can see, each symbol appears three times, because each edge lies on the boundary of three quadrilaterals. Since the scheme is not flat it is
nonorientable. Note that the $1$-skeleton, i.e. the
skeleton graph of the tesseract, is the $4$-hypercube graph, $Q_4$.
\hfill $\Diamond$
\end{example}

From now on, we will only consider flat polygonal complexes. Originally the term polygonal complex 
was reserved for flat polygonal complexes. See for instance chapter by Pisanski and Potočnik in  \cite{PiPo2004}. More information about maps can be obtained from \cite{GrYe2004}. The following example shows how one can distinguish between a tetrahedron 
and a tetrahedron with one face missing.

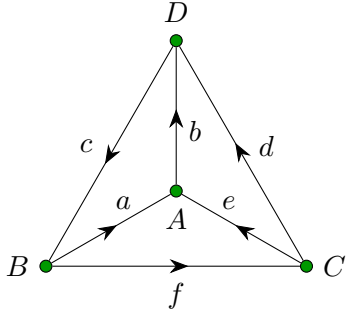
\begin{figure}[!htb]
\centering
\begin{tikzpicture}
\tikzstyle{every node}=[draw,circle,inner sep=1.6pt,fill=green!60!black] 
\tikzstyle{edge}=[draw,decoration={markings,mark=at position 0.55 with {\arrow{Stealth[scale=1.5]}}},postaction={decorate}] 
\node[label=-90:$A$] (a) at (0, 0) {};
\node[label=90:$D$] (d) at (90:2) {};
\node[label=0:$C$] (c) at (-30:2) {};
\node[label=180:$B$] (b) at (210:2) {};
\path[edge] (b) -- (c); 
\path[edge] (c) -- (d);
\path[edge] (d) -- (b);
\path[edge] (a) -- (d);
\path[edge] (b) -- (a);
\path[edge] (c) -- (a);
\node[draw=none,fill=none] at (0, -1.4) {$f$}; 
\node[draw=none,fill=none] at (1.2, 0.6) {$d$}; 
\node[draw=none,fill=none] at (-1.2, 0.6) {$c$}; 
\node[draw=none,fill=none] at (-0.7, -0.15) {$a$}; 
\node[draw=none,fill=none] at (0.7, -0.15) {$e$}; 
\node[draw=none,fill=none] at (0.25, 0.8) {$b$}; 
\end{tikzpicture}
\caption{Open and closed tetrahedra. The Schlegel diagram with the infinite face BCD included
represents the closed tetrahedron. In the open tetrahedron the final triangular
face $cfd$ ($BCD$) has been removed.}
\label{fig:tetrahedra}
\end{figure}

\begin{example}
Consider the scheme:
\begin{equation}
\Theta = \;
\begin{matrix*}[l]
a & b & c \\
a^- & f & e \\
b^- & e^- & f \\
f^- & c^- & d^-
\end{matrix*}
\end{equation}
This represents a tetrahedron. There are six edges and each row corresponds to a triangular face. All symbols in $\Theta$
appear twice, and hence the corresponding surface is closed (has no boundary). The surface in this case is a sphere. 
By removing a face, for instance the last one,  we obtain a connected scheme:
\begin{equation}
\Theta' = \;
\begin{matrix*}[l]
a & b & c \\
a^- & e & e \\
b^- & e^- & f 
\end{matrix*}
\end{equation}
that represents a tetrahedron with one face missing. The symbols $c,d,f$ each occur only once and the corresponding surface is a disk.
\hfill $\Diamond$
\end{example}

\begin{figure}[!htb]
\centering
\begin{minipage}[b]{\linewidth}
\centering
\begin{tikzpicture}[scale=1.0]
\tikzstyle{every node}=[draw,circle,inner sep=1.6pt,fill=green!60!black] 
\tikzstyle{edge}=[draw,decoration={markings,mark=at position 0.55 with {\arrow{Stealth[scale=1.5]}}},postaction={decorate}] 
\node[] (a) at (0, 0) {};
\node[] (b) at (1, 0) {};
\node[] (c) at (1, 1) {};
\node[] (d) at (0, 1) {};
\node[] (a2) at (-1, -1) {};
\node[] (b2) at (2, -1) {};
\node[] (c2) at (2, 2) {};
\node[] (d2) at (-1, 2) {};
\path[edge] (a) -- (b); 
\path[edge] (b) -- (c); 
\path[edge] (c) -- (d); 
\path[edge] (d) -- (a); 
\path[edge] (a2) -- (b2); 
\path[edge] (b2) -- (c2); 
\path[edge] (c2) -- (d2); 
\path[edge] (d2) -- (a2); 
\path[edge] (a2) -- (a); 
\path[edge] (b2) -- (b); 
\path[edge] (c2) -- (c); 
\path[edge] (d2) -- (d); 
\node[draw=none,fill=none] at (0.5, -1.2) {$a$}; 
\node[draw=none,fill=none] at (0.5, 0.2) {$e$}; 
\node[draw=none,fill=none] at (0.5, 1.2) {$g$}; 
\node[draw=none,fill=none] at (0.5, 2.2) {$c$}; 
\node[draw=none,fill=none] at (-1.3, 0.5) {$d$}; 
\node[draw=none,fill=none] at (-0.2, 0.5) {$h$}; 
\node[draw=none,fill=none] at (1.25, 0.5) {$f$}; 
\node[draw=none,fill=none] at (2.25, 0.5) {$b$}; 
\end{tikzpicture}
\subcaption{$\Sigma$}\label{fig:threeembeddings1}
\end{minipage}

\begin{minipage}[b]{0.45\linewidth}
\centering
\begin{tikzpicture}[scale=1.3]
\tikzstyle{every node}=[draw,circle,inner sep=1.6pt,fill=green!60!black] 
\tikzstyle{edge}=[draw,decoration={markings,mark=at position 0.6 with {\arrow{Stealth[scale=1.5]}}},postaction={decorate}] 
\tikzstyle{torus1}=[draw,decoration={markings,mark=at position 0.5 with {\arrow{>[scale=1.5]}},
,mark=at position 0.55 with {\arrow{>[scale=1.5]}}},postaction={decorate}] 
\tikzstyle{torus2}=[draw,decoration={markings,mark=at position 0.5 with {\arrow{>[scale=1.5]}}},postaction={decorate}] 
\node[] (a) at (0, 0) {};
\node[] (b) at (1, 0) {};
\node[] (c) at (1, 1) {};
\node[] (d) at (0, 1) {};
\node[] (a2) at (-1, 0) {};
\node[] (b2) at (2, 0) {};
\node[] (c2) at (2, 1) {};
\node[] (d2) at (-1, 1) {};
\path[edge] (a2) -- (a);
\path[edge] (a) -- (b);
\path[edge] (b) -- (b2);
\path[edge] (d2) -- (d);
\path[edge] (d) -- (c);
\path[edge] (c) -- (c2);
\path[edge] (c) -- (b);
\path[edge] (d2) -- (a2);
\path[edge] (-1.8, 0) -- (a2);
\path[edge] (b2) -- (2.8, 0);
\path[edge] (c2) -- (2.8, 1);
\path[edge] (-1.8, 1) -- (d2);
\path[edge] (d) -- (0, 1.7);
\path[edge] (0, -0.7) -- (a);
\path[edge] (c2) -- (2, 1.7);
\path[edge] (2, -0.7) -- (b2);
\path[torus1] (-1.8, -0.7) -- (-1.8, 1.7);
\path[torus1] (2.8, -0.7) -- (2.8, 1.7);
\path[torus2] (-1.8, -0.7) -- (2.8, -0.7);
\path[torus2] (-1.8, 1.7) -- (2.8, 1.7);
\node[draw=none,fill=none] at (2.5, 0.2) {$h$}; 
\node[draw=none,fill=none] at (1.5, 0.2) {$g$}; 
\node[draw=none,fill=none] at (0.5, 0.2) {$f$}; 
\node[draw=none,fill=none] at (-0.5, 0.2) {$e$}; 
\node[draw=none,fill=none] at (-1.5, 0.2) {$h$}; 
\node[draw=none,fill=none] at (2.5, 0.8) {$d$}; 
\node[draw=none,fill=none] at (1.5, 0.8) {$c$}; 
\node[draw=none,fill=none] at (0.5, 0.8) {$b$}; 
\node[draw=none,fill=none] at (-0.5, 0.8) {$a$}; 
\node[draw=none,fill=none] at (-1.5, 0.8) {$d$}; 
\node[draw=none,fill=none] at (1.2, 0.5) {$k$}; 
\node[draw=none,fill=none] at (-0.8, 0.5) {$i$}; 
\node[draw=none,fill=none] at (2.2, 1.5) {$l$}; 
\node[draw=none,fill=none] at (0.2, 1.5) {$j$}; 
\node[draw=none,fill=none] at (2.2, -0.5) {$l$}; 
\node[draw=none,fill=none] at (0.2, -0.5) {$j$}; 
\end{tikzpicture}
\subcaption{$\Sigma'$}\label{fig:threeembeddings2}
\end{minipage}
\begin{minipage}[b]{0.45\linewidth}
\centering
\begin{tikzpicture}[scale=1.3]
\tikzstyle{every node}=[draw,circle,inner sep=1.6pt,fill=green!60!black] 
\tikzstyle{edge}=[draw,decoration={markings,mark=at position 0.6 with {\arrow{Stealth[scale=1.5]}}},postaction={decorate}] 
\tikzstyle{torus1}=[draw,decoration={markings,mark=at position 0.5 with {\arrow{>[scale=1.5]}},
,mark=at position 0.55 with {\arrow{>[scale=1.5]}}},postaction={decorate}] 
\tikzstyle{torus2}=[draw,decoration={markings,mark=at position 0.5 with {\arrow{>[scale=1.5]}}},postaction={decorate}] 
\node[] (a) at (0, 0) {};
\node[] (b) at (1, 0) {};
\node[] (c) at (1, 1) {};
\node[] (d) at (0, 1) {};
\node[] (a2) at (-1, 0) {};
\node[] (b2) at (2, 0) {};
\node[] (c2) at (2, 1) {};
\node[] (d2) at (-1, 1) {};
\path[edge] (a) -- (a2);
\path[edge] (b) -- (a);
\path[edge] (b2) -- (b);
\path[edge] (d2) -- (d);
\path[edge] (d) -- (c);
\path[edge] (c) -- (c2);
\path[edge] (c2) -- (b);
\path[edge] (d) -- (a2);
\path[edge] (a2) -- (-1.8, 0);
\path[edge] (2.8, 0) -- (b2);
\path[edge] (c2) -- (2.8, 1);
\path[edge] (-1.8, 1) -- (d2);
\path[edge] (d2) -- (-0.5, 1.7);
\path[edge] (-0.5, -0.7) -- (a);
\path[edge] (c) -- (1.5, 1.7);
\path[edge] (1.5, -0.7) -- (b2);
\path[torus1] (-1.8, -0.7) -- (-1.8, 1.7);
\path[torus1] (2.8, -0.7) -- (2.8, 1.7);
\path[torus2] (-1.8, -0.7) -- (2.8, -0.7);
\path[torus2] (-1.8, 1.7) -- (2.8, 1.7);
\node[draw=none,fill=none] at (2.5, -0.2) {$e$}; 
\node[draw=none,fill=none] at (1.3, -0.2) {$f$}; 
\node[draw=none,fill=none] at (0.5, -0.2) {$g$}; 
\node[draw=none,fill=none] at (-0.7, -0.2) {$h$}; 
\node[draw=none,fill=none] at (-1.5, -0.2) {$e$}; 
\node[draw=none,fill=none] at (2.5, 0.8) {$c$}; 
\node[draw=none,fill=none] at (1.3, 0.8) {$b$}; 
\node[draw=none,fill=none] at (0.5, 0.8) {$a$}; 
\node[draw=none,fill=none] at (-0.7, 0.8) {$d$}; 
\node[draw=none,fill=none] at (-1.5, 0.8) {$c$}; 
\node[draw=none,fill=none] at (1.7, 0.5) {$k$}; 
\node[draw=none,fill=none] at (-0.3, 0.5) {$i$}; 
\node[draw=none,fill=none] at (1.6, 1.5) {$j$}; 
\node[draw=none,fill=none] at (-0.4, 1.5) {$l$}; 
\node[draw=none,fill=none] at (1.9, -0.5) {$j$}; 
\node[draw=none,fill=none] at (-0.1, -0.5) {$l$}; 
\end{tikzpicture}
\subcaption{$\Sigma''$}\label{fig:threeembeddings3}
\end{minipage}
\caption{Three embeddings of the cube graph $Q_3$.}
\label{fig:threeembeddings}
\end{figure}
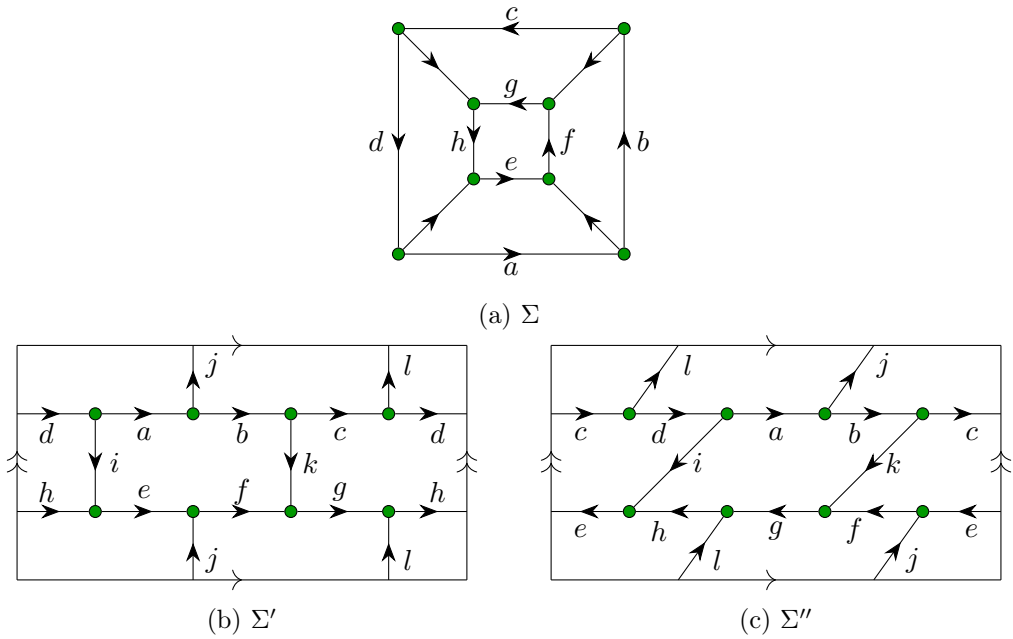

\begin{example}
Here are three maps that all have the same skeleton, namely the cube graph, $Q_3$:
\begin{equation}
\Sigma = \;
\begin{matrix*}[l]
a & b & c & d\\
h^- & g^- & f^- &  e^-  \\
a^- & i & e & j^-   \\
 j & f & k^- & b^-    \\
g &  l^- & c^- & k      \\
d^- &  l & h & i^- 
\end{matrix*}
\end{equation}

\begin{equation}
\Sigma' = \;
\begin{matrix}
a & b & k&f^-&e^-&i^-\\
c & d &i & h^- & g^- & k^-\\
f & g & l^- & c^- & b^- & j\\
h & e & j^- & a^- & d^- & l
\end{matrix}
\end{equation}

\begin{equation}
\Sigma'' = \;
\begin{matrix}
a & b & k&g&h&i^-\\
c & d &i & e & f & k^-\\
l & g^- & f^- & j^- & a^- & d^-  \\
b & c & l & h & e & j^-
\end{matrix}
\end{equation}
Note that $\Sigma$ describes the usual hexahedron, i.e.\ the surface of the cube.  $\Sigma'$ and $\Sigma''$ describe
two non-equivalent toroidal polyhexes. All three maps share the same underlying skeleton, the cube graph $Q_3$.
However, the embeddings $\Sigma'$ and $\Sigma''$ are clearly distinct. $\Sigma''$ has the property that each pair of faces intersects
in exactly two edges which are antipodal in each face, whereas $\Sigma'$ does not. $\Sigma''$ is a regular map \cite{Coxeter1973,McMullen2002},
a generalisation of Platonic polyhedra.

\hfill $\Diamond$
\end{example}

The above example raises an interesting question: Which toroidal polyhexes are completely determined by their skeleta?

\section{Equilinear catacondensed chemical hexagonal complexes}

Catacondensed chemical hexagonal complexes are characterized by the following rules: they are connected \ref{prop:connected}, flat \ref{prop:flat}, hexagonal \ref{prop:hexagonal}
and catacondensed \ref{prop:catacondensed}. These rules imply that such a complex is also chemical \ref{prop:chemical}. We may view such a complex as a collection of branching hexagons 
that are connected by chains  of hexagons. If each hexagonal chain is linear (property \ref{prop:linear} holds for the complex), we say that such a structure
is a \emph{linear catacondensed chemical hexagonal complex}. Moreover, if all linear hexagonal chains are of the same length, i.e.\ contain the same number of
hexagons, these structures are called \emph{equilinear}.  We denote by $l$ the common length of these linear chains.

\begin{figure}[!ht]
\centering
\includegraphics[scale=0.6]{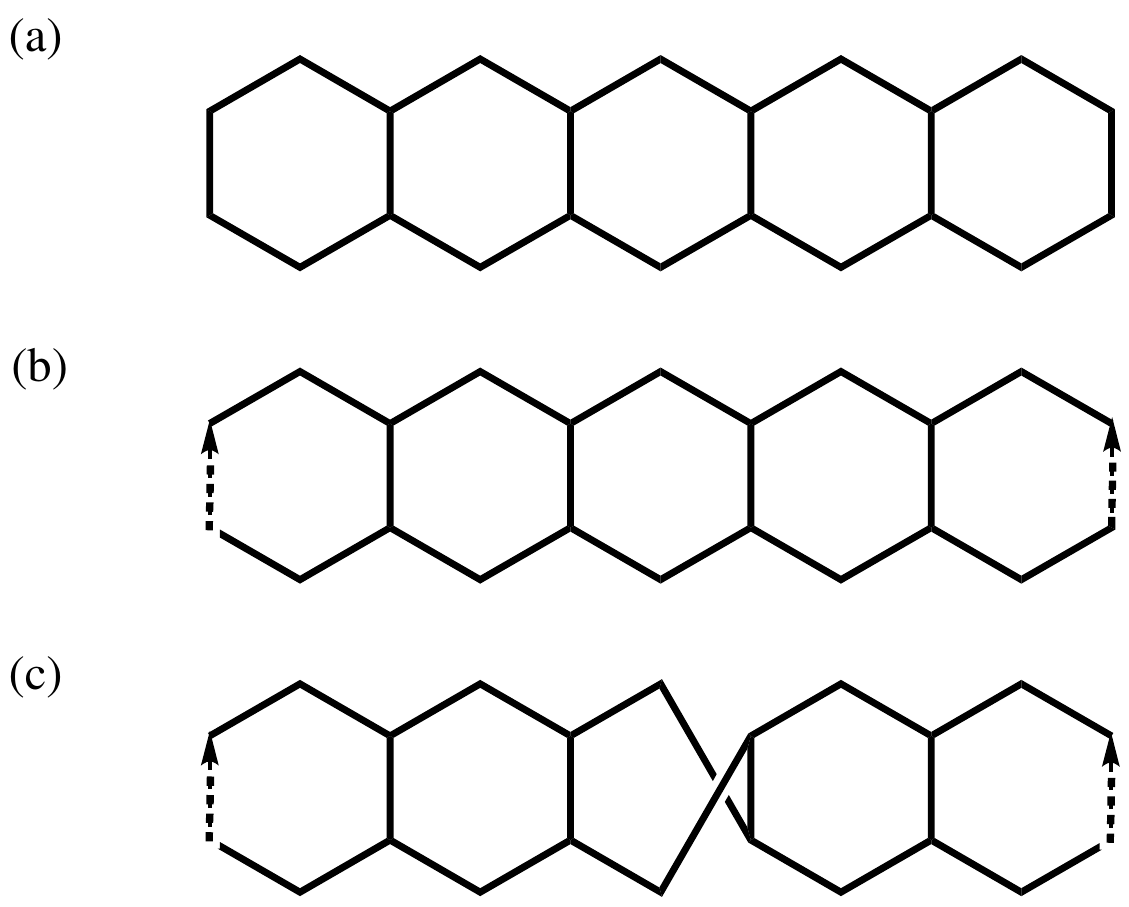}
\caption{Unbranched catacondensed chemical hexagonal complexes:
(a) linear polyacene $P_6$, (b) untwisted cyclacene $C_6$,
and (c) M{\"o}bius cyclacene $M_6$. For the two cyclacene
cases, the arrowed left and right edges are to be identified.}
\label{fig:thethreenonbranched}
\end{figure}

\begin{figure}[!htbp]
\centering
\includegraphics[scale=0.6]{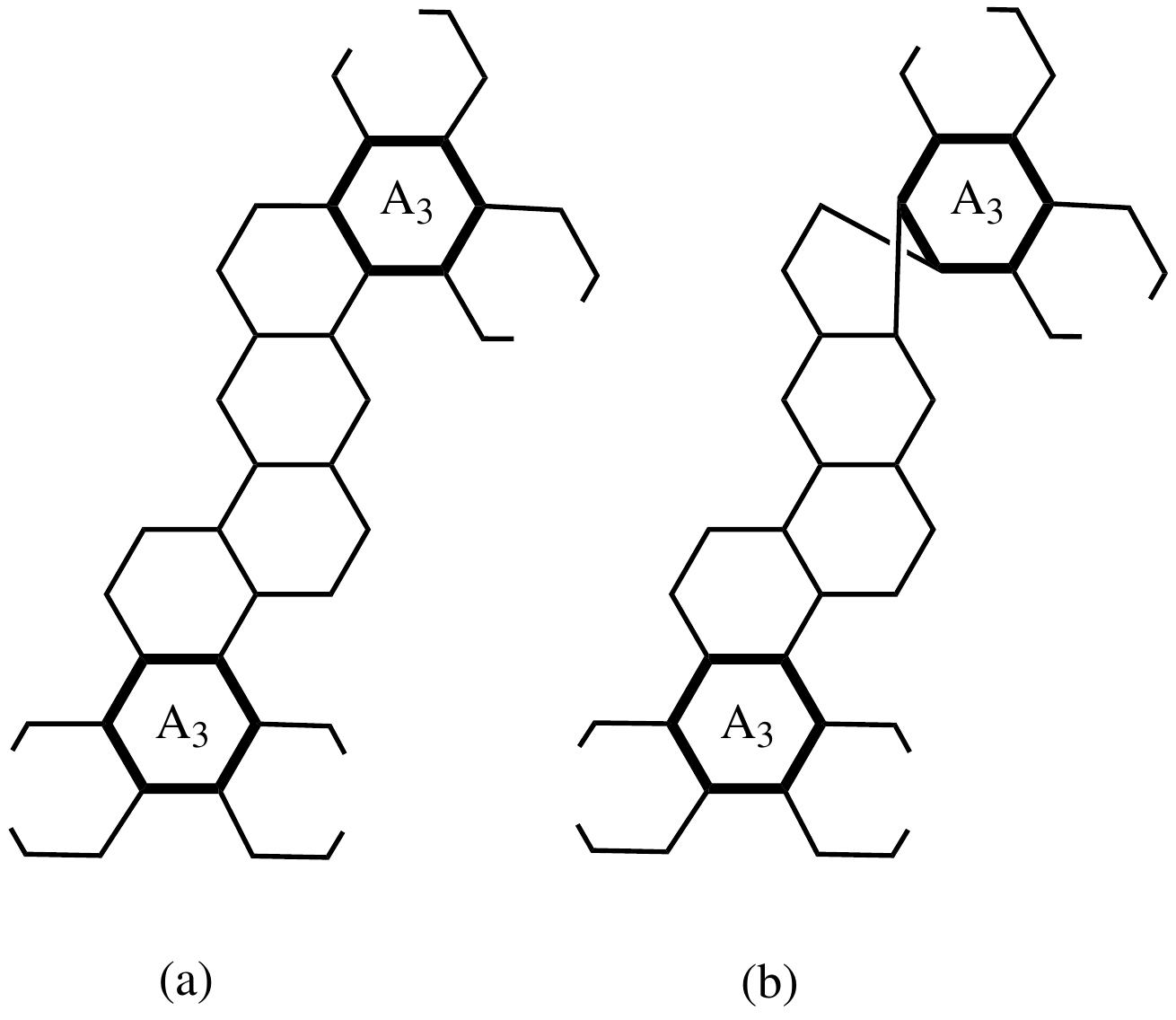}
\caption{Arc labelling in the blueprint map. In the untwisted (a) and the twisted (b) the word $w$ is $RLSR$ taken in the direction from bottom to top or $LSRL$ from 
top to bottom, whilst $\sigma = +$ and $\sigma = -$, respectively.} 
\label{fig:words}
\end{figure}

In the unbranched case (where \ref{prop:unbranched} holds), for a given $l$, there are only three catacondensed structures: $P_l, C_l$ and $M_l$, 
as illustrated in Figure~\ref{fig:thethreenonbranched}.
In the branched case, there are three types of hexagon: \emph{branching} (attached to three hexagons, i.e.\ of type $A_3$ in the notation of \cite{gutman1989}), 
\emph{connecting} (attached to two hexagons, i.e.\ of type $A_2$ or $L_2$), and \emph{terminal} (attached to a single hexagon, i.e.\ of type $L_1$). 
Such a structure defines a labelled 1-3 map (i.e.\ vertices are of degrees $1$ and $3$), called the \emph{blueprint map}.
A map is a graph together with a rotation projection \cite{Gross1987}.
The underlying graph is called the \emph{blueprint graph}. 
In the map, each arc is labelled by a pair $(w, \sigma)$, where $w$ is a word over the alphabet $\{L, R, S\}$ and $\sigma \in \{+, -\}$.
 The reverse of a label $(w, \sigma)$ is the label $(w^\rho, \sigma)$, where $w^\rho$ is the reverse of the word in which symbols $L$ and $R$ are interchanged.
 Labels are assigned to arcs (half-edges) and two opposite arcs are assigned reverse labels.
For instance, reverse of the label $(RLSR, +)$ is the label $(LSRL, +)$.
Degree $3$ vertices correspond to branching hexagons, while vertices of degree $1$ correspond to terminal hexagons. 
The connecting hexagons are implicitly described by the labels; see Figure~\ref{fig:words}.

In the equilinear case, the description given above can be simplified. Words in labels on arcs comprise only the letter $S$.
Therefore each word can be described by giving its length, and then only one integer parameter is needed.

\begin{figure}[!htb]
\centering
\begin{minipage}[b]{\linewidth}
\centering
\subcaption{Zero vertices of degree $3$:}\label{fig:blue0}
\begin{tikzpicture}[scale=0.7]
\tikzstyle{every node}=[draw,circle,inner sep=1.6pt,fill=green!60!black] 
\tikzstyle{edge}=[draw] 
\node (a) at (0, 1) {};
\node (b) at (0, 0) {};
\path[edge] (a) -- (b);
\end{tikzpicture}
$\mathcal{M}_{1,0}^{0}$ 
\hspace{1cm}
\begin{tikzpicture}[scale=0.7]
\tikzstyle{every node}=[draw,circle,inner sep=1.6pt,fill=green!60!black] 
\tikzstyle{edge}=[draw] 
\tikzstyle{medge}=[draw,decoration={markings,mark=at position 0.5 with {\draw(.15, .15) -- (-.15, -.15);\draw(-.15, .15) -- (.15, -.15);}},postaction={decorate}] 
\path[edge] (0, -1-0.6) circle (0.6);
\end{tikzpicture}
$\mathcal{M}_{2,0}^{0}$
\hspace{1cm}
\begin{tikzpicture}[scale=0.7]
\tikzstyle{every node}=[draw,circle,inner sep=1.6pt,fill=green!60!black] 
\tikzstyle{edge}=[draw] 
\tikzstyle{medge}=[draw,decoration={markings,mark=at position 0.5 with {\draw(.15, .15) -- (-.15, -.15);\draw(-.15, .15) -- (.15, -.15);}},postaction={decorate}] 
\path[medge] (0, -1-0.6) circle (0.6);
\end{tikzpicture}
$\mathcal{M}_{2,1}^{0}$
\end{minipage}

\vspace{\baselineskip}
\begin{minipage}[b]{\linewidth}
\centering
\subcaption{One vertex of degree $3$:}\label{fig:blue1}
\begin{tikzpicture}[scale=0.7]
\tikzstyle{every node}=[draw,circle,inner sep=1.6pt,fill=green!60!black] 
\tikzstyle{edge}=[draw] 
\tikzstyle{medge}=[draw,decoration={markings,mark=at position 0.25 with {\draw(.15, .15) -- (-.15, -.15);\draw(-.15, .15) -- (.15, -.15);}},postaction={decorate}] 
\tikzstyle{medge2}=[draw,decoration={markings,mark=at position 0.75 with {\draw(.15, .15) -- (-.15, -.15);\draw(-.15, .15) -- (.15, -.15);}},postaction={decorate}] 
\node (a) at (0, 1) {};
\node (b) at (0, 0) {};
\node (c1) at ($ (0, 1) + (30:1) $) {};
\node (c2) at ($ (0, 1) + (150:1) $) {};
\path[edge] (a) -- (b);
\path[edge] (c1) -- (a) -- (c2);
\end{tikzpicture}
$\mathcal{M}_{1,0}^{1}$ 
\hspace{1cm}
\begin{tikzpicture}[scale=0.7]
\tikzstyle{every node}=[draw,circle,inner sep=1.6pt,fill=green!60!black] 
\tikzstyle{edge}=[draw] 
\tikzstyle{medge}=[draw,decoration={markings,mark=at position 0.25 with {\draw(.15, .15) -- (-.15, -.15);\draw(-.15, .15) -- (.15, -.15);}},postaction={decorate}] 
\tikzstyle{medge2}=[draw,decoration={markings,mark=at position 0.75 with {\draw(.15, .15) -- (-.15, -.15);\draw(-.15, .15) -- (.15, -.15);}},postaction={decorate}] 
\path[edge] (0, 1+0.6) circle (0.6);
\node (a) at (0, 1) {};
\node (b) at (0, 0) {};
\path[edge] (a) -- (b);
\end{tikzpicture}
$\mathcal{M}_{2,0}^{1}$ 
\hspace{1cm}
\begin{tikzpicture}[scale=0.7]
\tikzstyle{every node}=[draw,circle,inner sep=1.6pt,fill=green!60!black] 
\tikzstyle{edge}=[draw] 
\tikzstyle{medge}=[draw,decoration={markings,mark=at position 0.25 with {\draw(.15, .15) -- (-.15, -.15);\draw(-.15, .15) -- (.15, -.15);}},postaction={decorate}] 
\tikzstyle{medge2}=[draw,decoration={markings,mark=at position 0.75 with {\draw(.15, .15) -- (-.15, -.15);\draw(-.15, .15) -- (.15, -.15);}},postaction={decorate}] 
\path[medge] (0, 1+0.6) circle (0.6);
\node (a) at (0, 1) {};
\node (b) at (0, 0) {};
\path[edge] (a) -- (b);
\end{tikzpicture}
$\mathcal{M}_{2,1}^{1}$ 
\end{minipage}

\vspace{\baselineskip}
\begin{minipage}[b]{\linewidth}
\centering
\subcaption{Two vertices of degree $3$, one connecting edge:}\label{fig:blue21}
\begin{tikzpicture}[scale=0.7]
\useasboundingbox (-0.7,-2.2-0.15) rectangle (0.7,+2.2+0.15);
\tikzstyle{every node}=[draw,circle,inner sep=1.6pt,fill=green!60!black] 
\tikzstyle{edge}=[draw] 
\tikzstyle{medge}=[draw,decoration={markings,mark=at position 0.25 with {\draw(.15, .15) -- (-.15, -.15);\draw(-.15, .15) -- (.15, -.15);}},postaction={decorate}] 
\node (a) at (0, 1) {};
\node (b) at (0, -1) {};
\path[edge] (a) -- (b);
\node (b1) at (-0.5, -2) {};
\node (b2) at (0.5, -2) {};
\path[edge] (b) -- (b1);
\path[edge] (b) -- (b2);
\node (a1) at (-0.5, 2) {};
\node (a2) at (0.5, 2) {};
\path[edge] (a) -- (a1);
\path[edge] (a) -- (a2);
\end{tikzpicture}
$\mathcal{M}_{1,0}^{2}$ 
\hspace{0.5cm}
\begin{tikzpicture}[scale=0.7]
\useasboundingbox (-0.7,-2.2-0.15) rectangle (0.7,+2.2+0.15);
\tikzstyle{every node}=[draw,circle,inner sep=1.6pt,fill=green!60!black] 
\tikzstyle{edge}=[draw] 
\tikzstyle{medge}=[draw,decoration={markings,mark=at position 0.25 with {\draw(.15, .15) -- (-.15, -.15);\draw(-.15, .15) -- (.15, -.15);}},postaction={decorate}] 
\path[edge] (0, 1+0.6) circle (0.6);
\node (a) at (0, 1) {};
\node (b) at (0, -1) {};
\path[edge] (a) -- (b);
\node (b1) at (-0.5, -2) {};
\node (b2) at (0.5, -2) {};
\path[edge] (b) -- (b1);
\path[edge] (b) -- (b2);
\end{tikzpicture}
$\mathcal{M}_{2,0}^{2}$ 
\hspace{0.5cm}
\begin{tikzpicture}[scale=0.7]
\useasboundingbox (-0.7,-2.2-0.15) rectangle (0.7,+2.2+0.15);
\tikzstyle{every node}=[draw,circle,inner sep=1.6pt,fill=green!60!black] 
\tikzstyle{edge}=[draw] 
\tikzstyle{medge}=[draw,decoration={markings,mark=at position 0.25 with {\draw(.15, .15) -- (-.15, -.15);\draw(-.15, .15) -- (.15, -.15);}},postaction={decorate}] 
\path[medge] (0, 1+0.6) circle (0.6);
\node (a) at (0, 1) {};
\node (b) at (0, -1) {};
\path[edge] (a) -- (b);
\node (b1) at (-0.5, -2) {};
\node (b2) at (0.5, -2) {};
\path[edge] (b) -- (b1);
\path[edge] (b) -- (b2);
\end{tikzpicture}
$\mathcal{M}_{2,1}^{2}$ 
\hspace{0.5cm}
\begin{tikzpicture}[scale=0.7]
\useasboundingbox (-0.7,-2.2-0.15) rectangle (0.7,+2.2+0.15);
\tikzstyle{every node}=[draw,circle,inner sep=1.6pt,fill=green!60!black] 
\tikzstyle{edge}=[draw] 
\tikzstyle{medge}=[draw,decoration={markings,mark=at position 0.25 with {\draw(.15, .15) -- (-.15, -.15);\draw(-.15, .15) -- (.15, -.15);}},postaction={decorate}] 
\path[edge] (0, 1+0.6) circle (0.6);
\path[edge] (0, -1-0.6) circle (0.6);
\node (a) at (0, 1) {};
\node (b) at (0, -1) {};
\path[edge] (a) -- (b);
\end{tikzpicture}
$\mathcal{M}_{3,0}^{2}$ 
\hspace{0.5cm}
\begin{tikzpicture}[scale=0.7,]
\useasboundingbox (-0.7,-2.2-0.15) rectangle (0.7,+2.2+0.15);
\tikzstyle{every node}=[draw,circle,inner sep=1.6pt,fill=green!60!black] 
\tikzstyle{edge}=[draw] 
\tikzstyle{medge}=[draw,decoration={markings,mark=at position 0.25 with {\draw(.15, .15) -- (-.15, -.15);\draw(-.15, .15) -- (.15, -.15);}},postaction={decorate}] 
\path[medge] (0, 1+0.6) circle (0.6);
\path[edge] (0, -1-0.6) circle (0.6);
\node (a) at (0, 1) {};
\node (b) at (0, -1) {};
\path[edge] (a) -- (b);
\end{tikzpicture}
$\mathcal{M}_{3,1}^{2}$ 
\hspace{0.5cm}
\begin{tikzpicture}[scale=0.7]
\useasboundingbox (-0.7,-2.2-0.15) rectangle (0.7,+2.2+0.15);
\tikzstyle{every node}=[draw,circle,inner sep=1.6pt,fill=green!60!black] 
\tikzstyle{edge}=[draw] 
\tikzstyle{medge}=[draw,decoration={markings,mark=at position 0.25 with {\draw(.15, .15) -- (-.15, -.15);\draw(-.15, .15) -- (.15, -.15);}},postaction={decorate}] 
\tikzstyle{medge2}=[draw,decoration={markings,mark=at position 0.75 with {\draw(.15, .15) -- (-.15, -.15);\draw(-.15, .15) -- (.15, -.15);}},postaction={decorate}] 
\path[medge] (0, 1+0.6) circle (0.6);
\path[medge2] (0, -1-0.6) circle (0.6);
\node (a) at (0, 1) {};
\node (b) at (0, -1) {};
\path[edge] (a) -- (b);
\end{tikzpicture}
$\mathcal{M}_{3,2}^{2}$ 
\end{minipage}

\vspace{\baselineskip}
\begin{minipage}[b]{\linewidth}
\centering
\subcaption{Two vertices of degree $3$, two connecting edges:}\label{fig:blue22}
\begin{tikzpicture}[scale=0.7]
\useasboundingbox (-0.7,-2-0.15) rectangle (0.7,+2+0.15);
\tikzstyle{every node}=[draw,circle,inner sep=1.6pt,fill=green!60!black] 
\tikzstyle{edge}=[draw] 
\tikzstyle{medge}=[draw,decoration={markings,mark=at position 0.5 with {\draw(.15, .15) -- (-.15, -.15);\draw(-.15, .15) -- (.15, -.15);}},postaction={decorate}] 
\node (a) at (0, 1) {};
\node (a2) at (0, 2) {};
\node (b) at (0, -1) {};
\node (b2) at (0, -2) {};
\path[edge] (a) -- (a2);
\path[edge] (b) -- (b2);
\path[edge] (a) .. controls (-0.7, 0.5) and (-0.7, -0.5) .. (b);
\path[edge] (a) .. controls (0.7, 0.5) and (0.7, -0.5) .. (b);
\end{tikzpicture}
$\mathcal{M}_{4,0}^{2}$ 
\hspace{1cm}
\begin{tikzpicture}[scale=0.7]
\useasboundingbox (-0.7,-2-0.15) rectangle (0.7,+2+0.15);
\tikzstyle{every node}=[draw,circle,inner sep=1.6pt,fill=green!60!black] 
\tikzstyle{edge}=[draw] 
\tikzstyle{medge}=[draw,decoration={markings,mark=at position 0.5 with {\draw(.15, .15) -- (-.15, -.15);\draw(-.15, .15) -- (.15, -.15);}},postaction={decorate}] 
\node (a) at (0, 1) {};
\node (a2) at (0, 2) {};
\node (b) at (0, -1) {};
\node (b2) at (0, -2) {};
\path[edge] (a) -- (a2);
\path[edge] (b) -- (b2);
\path[medge] (a) .. controls (-0.7, 0.5) and (-0.7, -0.5) .. (b);
\path[edge] (a) .. controls (0.7, 0.5) and (0.7, -0.5) .. (b);
\end{tikzpicture}
$\mathcal{M}_{4,1}^{2}$ 
\hspace{1cm}
\begin{tikzpicture}[scale=0.7]
\useasboundingbox (-0.7,-2-0.15) rectangle (0.7,+2+0.15);
\tikzstyle{every node}=[draw,circle,inner sep=1.6pt,fill=green!60!black] 
\tikzstyle{edge}=[draw] 
\tikzstyle{medge}=[draw,decoration={markings,mark=at position 0.5 with {\draw(.15, .15) -- (-.15, -.15);\draw(-.15, .15) -- (.15, -.15);}},postaction={decorate}] 
\node (a) at (0, 1) {};
\node (a2) at (0, 2) {};
\node (b) at (0, -1) {};
\node (b2) at (0, 0) {};
\path[edge] (a) -- (a2);
\path[edge] (b) -- (b2);
\path[edge] (a) .. controls (-0.7, 0.5) and (-0.7, -0.5) .. (b);
\path[edge] (a) .. controls (0.7, 0.5) and (0.7, -0.5) .. (b);
\end{tikzpicture}
$\mathcal{M}_{5,0}^{2}$ 
\hspace{1cm}
\begin{tikzpicture}[scale=0.7]
\useasboundingbox (-0.7,-2-0.15) rectangle (0.7,+2+0.15);
\tikzstyle{every node}=[draw,circle,inner sep=1.6pt,fill=green!60!black] 
\tikzstyle{edge}=[draw] 
\tikzstyle{medge}=[draw,decoration={markings,mark=at position 0.5 with {\draw(.15, .15) -- (-.15, -.15);\draw(-.15, .15) -- (.15, -.15);}},postaction={decorate}] 
\node (a) at (0, 1) {};
\node (a2) at (0, 2) {};
\node (b) at (0, -1) {};
\node (b2) at (0, 0) {};
\path[edge] (a) -- (a2);
\path[edge] (b) -- (b2);
\path[medge] (a) .. controls (-0.7, 0.5) and (-0.7, -0.5) .. (b);
\path[edge] (a) .. controls (0.7, 0.5) and (0.7, -0.5) .. (b);
\end{tikzpicture}
$\mathcal{M}_{5,1}^{2}$ 
\end{minipage}

\vspace{\baselineskip}
\begin{minipage}[b]{\linewidth}
\centering
\subcaption{Two vertices of degree $3$, three connecting edges:}\label{fig:blue22}
\begin{tikzpicture}[scale=1]
\useasboundingbox (-0.7,-1-0.15) rectangle (0.7,+1+0.15);
\tikzstyle{every node}=[draw,circle,inner sep=1.6pt,fill=green!60!black] 
\tikzstyle{edge}=[draw] 
\tikzstyle{medge}=[draw,decoration={markings,mark=at position 0.6 with {\arrow{Stealth[scale=1.5]}}},postaction={decorate}] 
\node (a) at (0, 1) {};
\node (b) at (0, -1) {};
\path[edge] (a) -- (b);
\path[edge] (a) .. controls (-0.7, 0.5) and (-0.7, -0.5) .. (b);
\path[edge] (a) .. controls (0.7, 0.5) and (0.7, -0.5) .. (b);
\end{tikzpicture}
$\mathcal{M}_{6,0}^{2}$ 
\hspace{1cm}
\begin{tikzpicture}[scale=1]
\useasboundingbox (-0.7,-1-0.15) rectangle (0.7,+1+0.15);
\tikzstyle{every node}=[draw,circle,inner sep=1.6pt,fill=green!60!black] 
\tikzstyle{edge}=[draw] 
\tikzstyle{medge}=[draw,decoration={markings,mark=at position 0.5 with {\draw(.15, .15) -- (-.15, -.15);\draw(-.15, .15) -- (.15, -.15);}},postaction={decorate}] 
\node (a) at (0, 1) {};
\node (b) at (0, -1) {};
\path[edge] (a) -- (b);
\path[medge] (a) .. controls (-0.7, 0.5) and (-0.7, -0.5) .. (b);
\path[edge] (a) .. controls (0.7, 0.5) and (0.7, -0.5) .. (b);
\end{tikzpicture}
$\mathcal{M}_{6,1}^{2}$ 
\hspace{1cm}
\begin{tikzpicture}[scale=1]
\useasboundingbox (-0.7,-1-0.15) rectangle (0.7,+1+0.15);
\tikzstyle{every node}=[draw,circle,inner sep=1.6pt,fill=green!60!black] 
\tikzstyle{edge}=[draw] 
\tikzstyle{medge}=[draw,decoration={markings,mark=at position 0.5 with {\draw(1, 1) -- (-1, -1);}},postaction={decorate}] 
\node (a) at (0, 1) {};
\node (b) at (0, -1) {};
\path[edge] (a) .. controls (-0.7, 0.5) and (-0.7, -0.5) .. (b);
\path[name path=p1] (a) .. controls (1.2, 0) and (0.0, 0.2) .. (b);
\path[name path=p2] (b) .. controls (1.2, 0) and (0.0, -0.2) .. (a);
\path [name intersections={of=p1 and p2,by=x}];
\path[edge] (a) .. controls (1.2, 0) and (0.0, 0.2) .. (b);
\node[circle,fill=white,draw=none,inner sep=2pt] at (x) {};
\path[edge] (b) .. controls (1.2, 0) and (0.0, -0.2) .. (a);
\end{tikzpicture}
$\mathcal{M}_{7,0}^{2}$ 
\hspace{1cm}
\begin{tikzpicture}[scale=1]
\useasboundingbox (-0.7,-1-0.15) rectangle (0.7,+1+0.15);
\tikzstyle{every node}=[draw,circle,inner sep=1.6pt,fill=green!60!black] 
\tikzstyle{edge}=[draw] 
\tikzstyle{medge}=[draw,decoration={markings,mark=at position 0.5 with {\draw(.15, .15) -- (-.15, -.15);\draw(-.15, .15) -- (.15, -.15);}},postaction={decorate}] 
\node (a) at (0, 1) {};
\node (b) at (0, -1) {};
\path[medge] (a) .. controls (-0.7, 0.5) and (-0.7, -0.5) .. (b);
\path[name path=p1] (a) .. controls (1.2, 0) and (0.0, 0.2) .. (b);
\path[name path=p2] (b) .. controls (1.2, 0) and (0.0, -0.2) .. (a);
\path [name intersections={of=p1 and p2,by=x}];
\path[edge] (a) .. controls (1.2, 0) and (0.0, 0.2) .. (b);
\node[circle,fill=white,draw=none,inner sep=2pt] at (x) {};
\path[edge] (b) .. controls (1.2, 0) and (0.0, -0.2) .. (a);
\end{tikzpicture}
$\mathcal{M}_{7,1}^{2}$ 
\end{minipage}

\caption{The blueprint maps with $n \leq 2$ trivalent vertices. The symbol
\tikzCross
on an edge represents a half-twist.
In maps $\mathcal{M}_{7,0}^{2}$ and $\mathcal{M}_{7,1}^{2}$ the crossing
edges are necessary, because the order of edges around a vertex is significant
and cannot be represented in a planar drawing.
}
\label{fig:blue}
\end{figure}
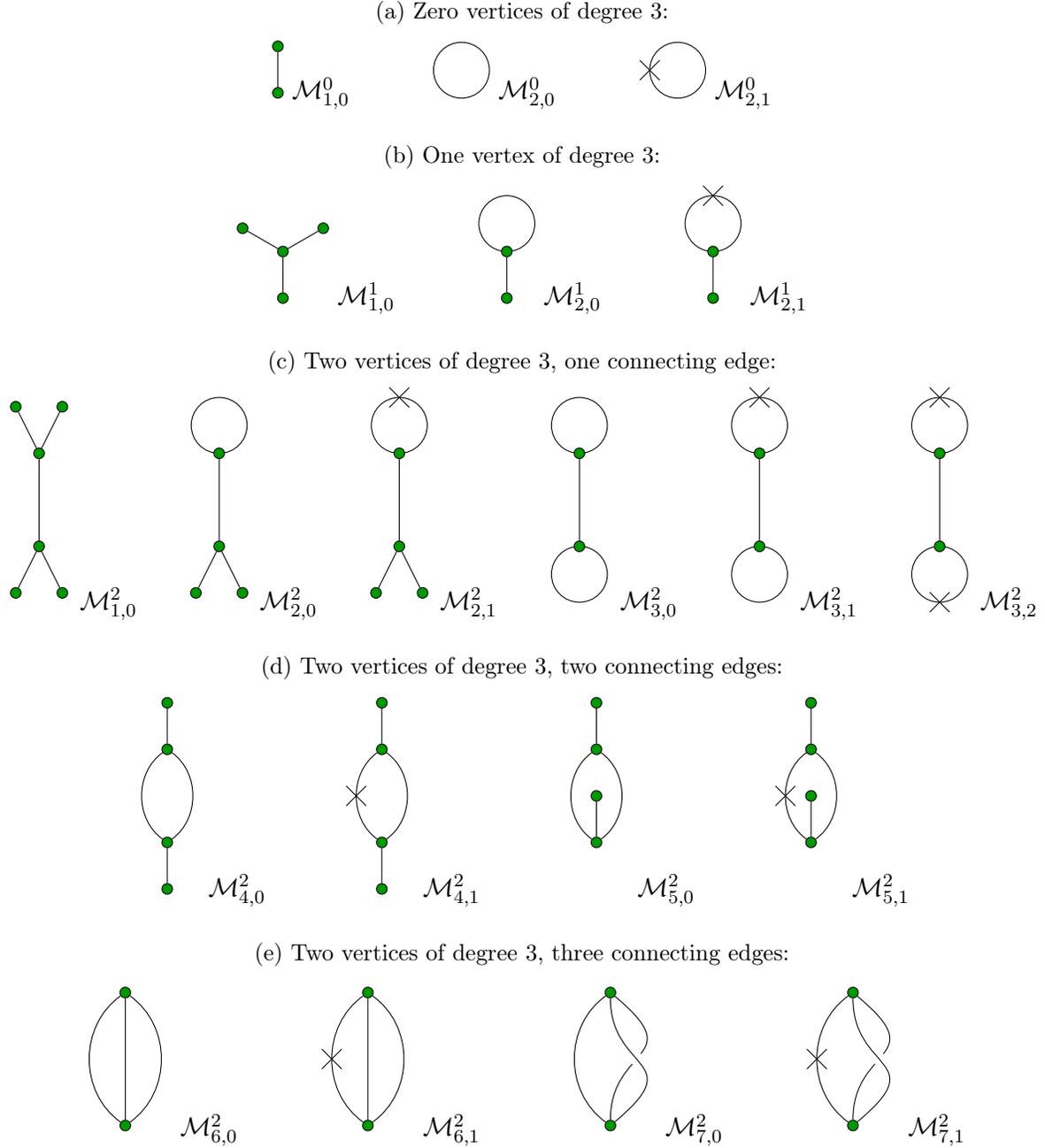

Figure~\ref{fig:blue} shows rotation projections for all blueprint maps on up to two trivalent vertices.
Note that in the case of zero vertices of degree $3$, two cases are anomalous, as they are free loops with no vertices at all.

\section{Kekul\'{e} structures in catacondensed coronoid complexes}

A natural question to ask is which flat hexagonal complexes admit a Kekul\'{e} structure. 
It is well known that all catacondensed benzenoids are Kekulean
\cite{gutman1989}.

\begin{theorem}
Any catacondensed flat hexagonal complex $B$  is Kekulean.
\end{theorem}

\noindent
Note that any catacondensed flat hexagonal complex is also a chemical hexagonal complex.

\begin{proof}
The proof is constructive -- we construct a perfect matching $\mathcal{M}$.
A catacondensed flat hexagonal complex contains only hexagons of type $L_1$, $L_2$, $A_2$ and $A_3$ (see \cite[p.~21]{gutman1989}).
In the first step we remove all hexagons of type $A_3$ from $B$. Edges labelled $a, b$ and $c$ in Figure~\ref{fig:types} will be
single bonds (they are not in the matching $\mathcal{M}$). By removing a hexagon of type $A_3$ we mean deleting edges $a$, $b$ and $c$.
\begin{figure}[!h]
\centering
\includegraphics[scale=0.6]{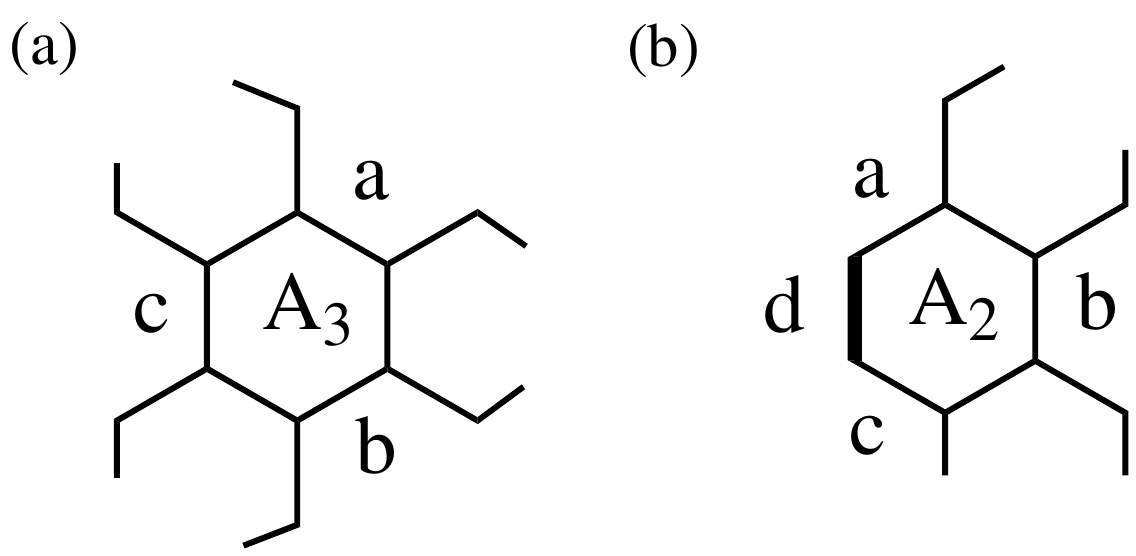}
\caption{Types of hexagons in catafused flat hexagonal complex $B$.}
\label{fig:types}
\end{figure}
In the second step we remove all hexagons of type $A_2$ from $B$. Edges labelled $a, b$ and $c$ in Figure~\ref{fig:types} will be
single bonds, whilst the edge $d$ will be double (we add it to $M$).
\begin{figure}[!h]
\centering
\includegraphics[scale=0.6]{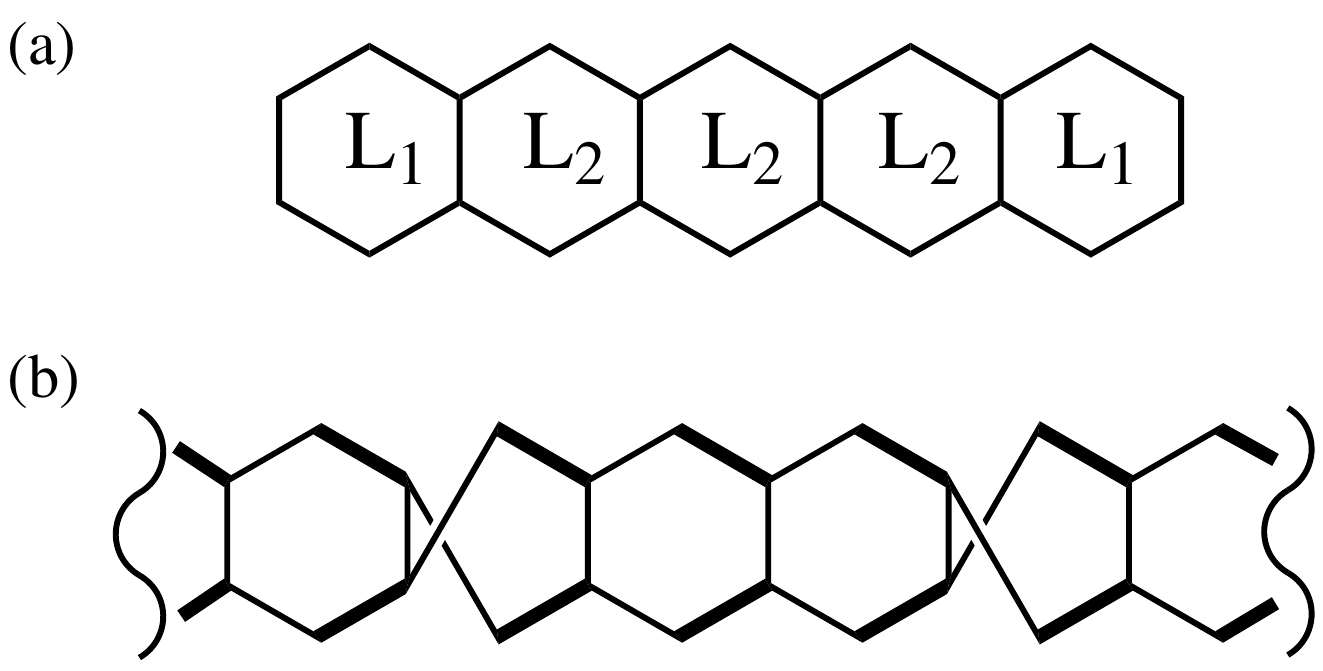}
\caption{The complex $B$ after deletion of hexagons of types $A_3$ and $A_2$.}
\label{fig:leftovers}
\end{figure}

What remains is a disjoint union of $k$ linear polyacene chains $P^i$ (see Figure~\ref{fig:leftovers}(a)) which may be of different lengths, 
$p$ untwisted cyclacenes $C^i$ (see Figure~\ref{fig:leftovers}(a)), $q$ twisted cyclacenes $T^i$ (see Figure~\ref{fig:leftovers}(b)),
and $m$ isolated $K_2$ fragments:
$$
\{P^1, P^2, \ldots, P^k\} \cup \{C^1, C^2, \ldots, C^p \} \cup \{ M^1, M^2, \ldots, M^q \} \cup  \{ K_2^1, K_2^2, \ldots, K_2^r \}.
$$
We add all isolated $K_2$ fragments to the perfect matching $\mathcal{M}$.  
All linear chains and cyclacenes are Kekulean. For each chain we may pick any of $l_i + 1$ perfect matchings. An untwisted cyclacene has $4$ perfect matchings,
and a twisted cyclacene has $2$ perfect matchings, hence
$
K(B) \geq 4^p 2^q \prod_{i=1}^{k} (l_i + 1).
$
\end{proof}

\begin{figure}[!b]
\centering
\includegraphics[width=0.9\textwidth]{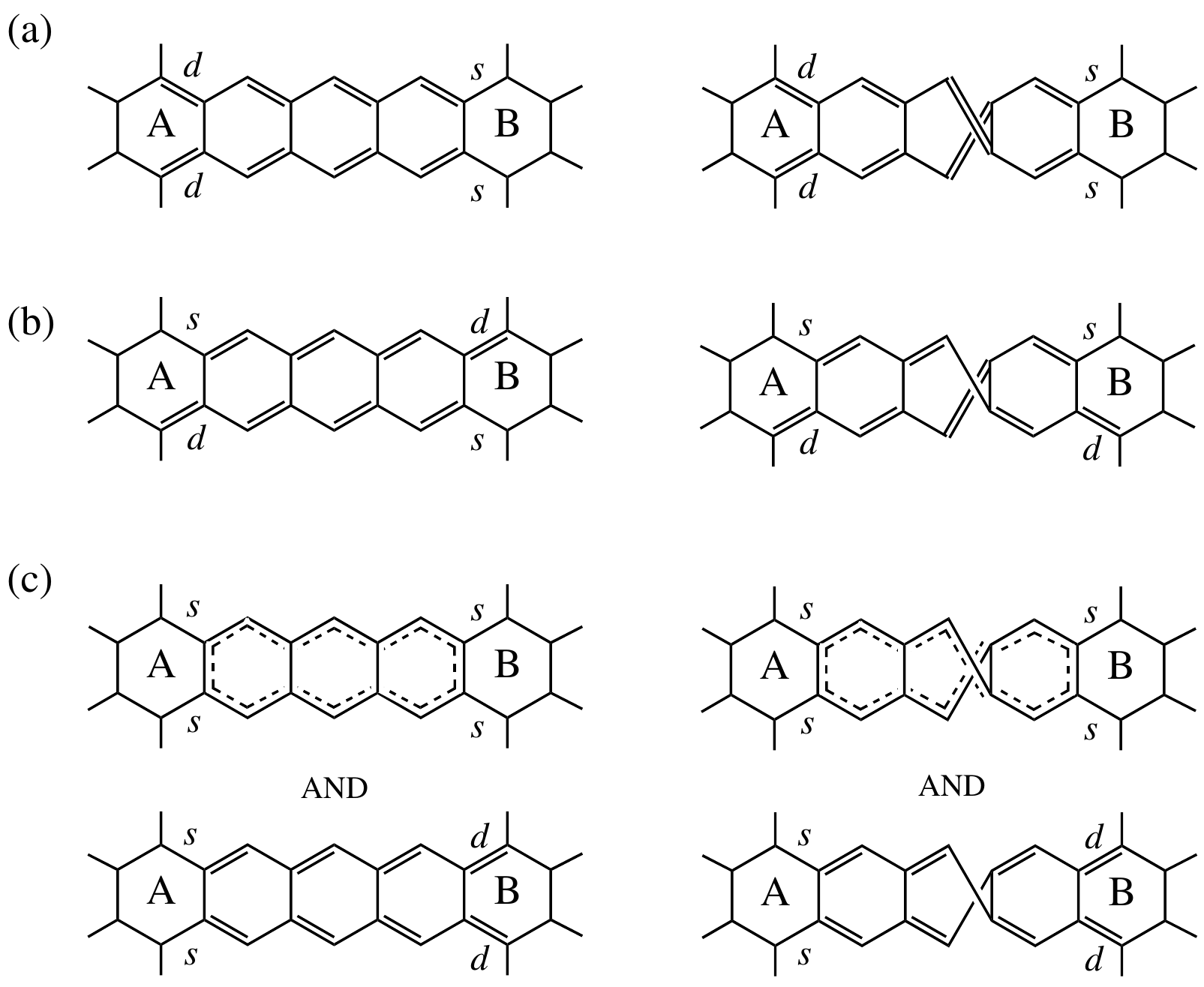}
\caption{Basic Rules for perfect matchings of hexagonal complexes derived with linear polyacene strips (illustrated for strips of length $3$).  They are: (a) the \textit{linear forcing} rule; (b) the {\it crossover} rule; (c) the {\it pairing rule}.  Hexagons A and B are derived from cubic vertices.  Fixing the illustrated {\it endo} bonds of hexagon A (denoted $d$ or $s$ for double or single) either forces (rules (a) and (b)) or rules out ((c)) given pairings of the corresponding {\it endo} bonds in hexagon B.  The panels show (left) the untwisted chain and (right) the chain with a M{\"o}bius half-twist.}
\label{fig:PWF4}
\end{figure}

This reasoning can be used as the basis of a simple procedure for counting the perfect matchings for a flat hexagonal complex 
with a given rotation scheme. For the linear polyacene motif, this leads to polynomial functions in $l + 1$. 
The detailed form of these expressions, the powers that appear and the coefficients that multiply them can be rationalised by thinking about a set of forcing rules
for replacements of edges of the cubic graph by linear chains of polyacenes.
In this case, three rules apply to allowed combinations of pairs of edges in the two branching hexagons (see Figure~\ref{fig:PWF4}). Rule (a) is the \emph{linear forcing} rule, by which two double bonds in A \emph{force} a fixed matching in the chain and two single bonds in B. Rule (b) is the \emph{crossover rule}, by which a single/double pair in A \emph{forces} a fixed matching in the chain and a double/single pair in B. Rule (c) is the \emph{pairing rule}, by which a pair of single bonds in A is \emph{compatible with} either a pair of single bonds or a pair of double bonds in B.
The pair of single bonds in B results from taking any of the $(l+1)$ perfect matchings of the intervening hexagons. The pair of double bonds of B arises from 
reversal of the linear forcing rule. (A single/double pair in B is ruled out by the crossover rule.)

\begin{figure}[!htbp]
\centering
\includegraphics[scale=0.6]{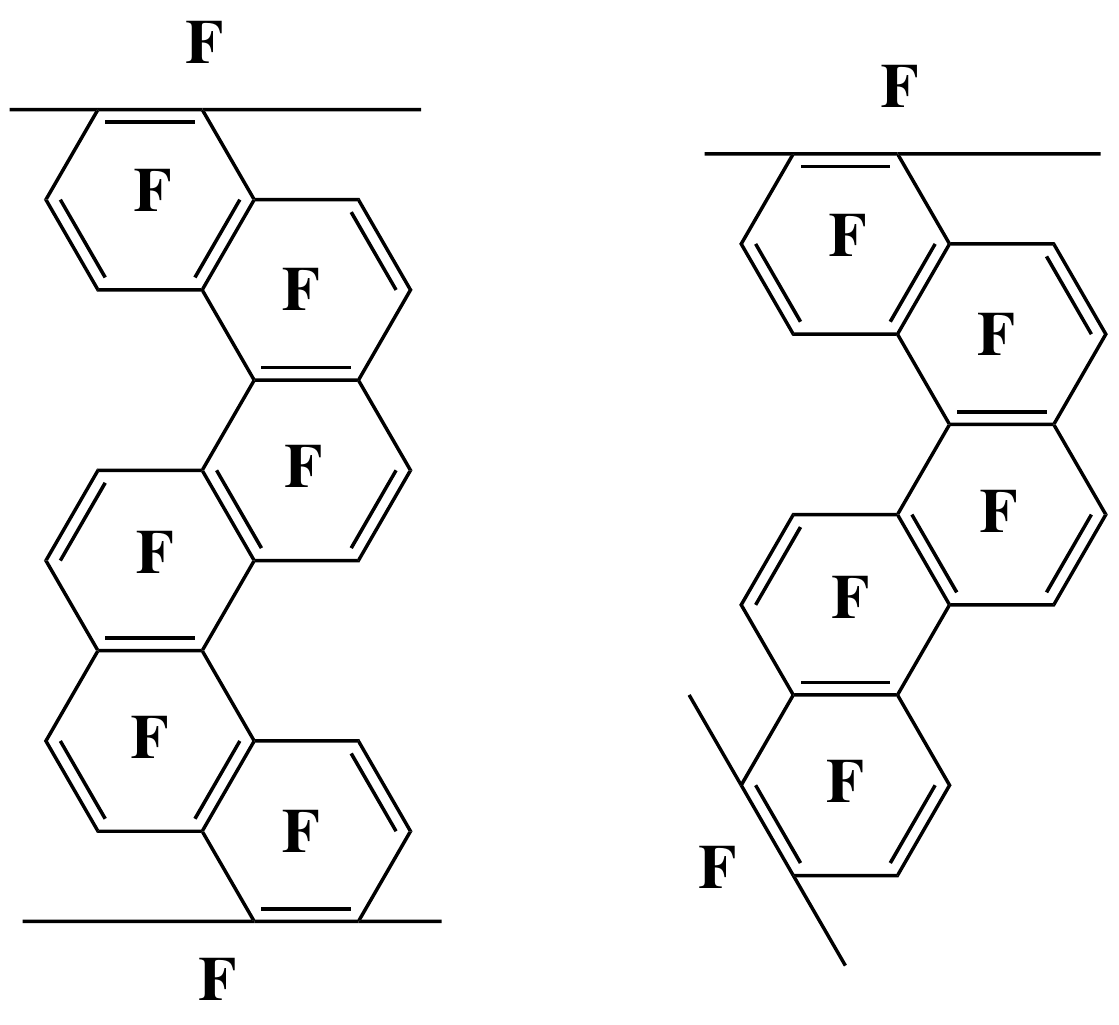}
\caption{A fully Fries hexagonal complex, i.e.\ one in which every hexagonal face includes three matched edges, can be constructed by 
using either an odd or or an even zig-zag fibonacene to inflate all edges of a cubic graph. 
Several attachment isomers are possible: for example, a fully Fries attachment isomer could be built using 
any external double bond in each terminal hexagon.
}
\label{fig:PWF7}
\end{figure}

Note that if we make a complex from a cubic graph with $m$ edges by using a straight chain of length $l$ on every edge, there is a term 
$(l+1)^m$ in the Kekul\'{e} count. Kinks in the chains will increase this leading term \cite{Balaban1989,Fowler2014}. 
E.g.\ fibonacene chains of length $l$ would lead to a term  $(F_{l+2})^m$, where $F_{l+2}$ is the $(l+2)$-th Fibonacci number. 
Fibonacene chains also allow favourable perfect matchings in which there are many hexagonal rings containing three double bonds, thus 
conforming to classical models of stability based on the ideas of Fries \cite{Fries1927} and Clar \cite{Clar1972}; see Figure~\ref{fig:PWF7}. 

Some explicit formulas for Kekul{\' e} counts of complexes built from cubic graphs and linear polyacenes are:

\begin{figure}[!htb]
\begin{tikzpicture}[scale=1]
\useasboundingbox (-1.5,-0.9) rectangle (1.5,1.6);
\tikzstyle{every node}=[draw,circle,inner sep=1.6pt,fill=green!60!black] 
\tikzstyle{edge}=[draw] 
\tikzstyle{medge}=[draw,decoration={markings,mark=at position 0.5 with {\draw(.15, .15) -- (-.15, -.15);\draw(-.15, .15) -- (.15, -.15);}},postaction={decorate}] 
\node (s) at (0, 0) {};
\node (b) at (90:1.5) {};
\node (a) at (-30:1.5) {};
\node (c) at (210:1.5) {};
\path[edge] (a) -- (s);
\path[edge] (a) -- (c);
\path[edge] (s) -- (b);
\path[edge] (s) -- (c);
\path[edge] (a) -- (b);
\path[edge] (b) -- (c);
\end{tikzpicture}$\mathcal{T}_{0,0}$
\begin{tikzpicture}[scale=1]
\useasboundingbox (-1.5,-0.9) rectangle (1.5,1.6);
\tikzstyle{every node}=[draw,circle,inner sep=1.6pt,fill=green!60!black] 
\tikzstyle{edge}=[draw] 
\tikzstyle{medge}=[draw,decoration={markings,mark=at position 0.5 with {\draw(.15, .15) -- (-.15, -.15);\draw(-.15, .15) -- (.15, -.15);}},postaction={decorate}] 
\node (s) at (0, 0) {};
\node (b) at (90:1.5) {};
\node (a) at (-30:1.5) {};
\node (c) at (210:1.5) {};
\path[edge] (a) -- (s);
\path[edge] (a) -- (c);
\path[edge] (s) -- (b);
\path[edge] (s) -- (c);
\path[edge] (a) -- (b);
\path[medge] (b) -- (c);
\end{tikzpicture}$\mathcal{T}_{0,1}$
\begin{tikzpicture}[scale=1]
\useasboundingbox (-1.5,-0.9) rectangle (1.5,1.6);
\tikzstyle{every node}=[draw,circle,inner sep=1.6pt,fill=green!60!black] 
\tikzstyle{edge}=[draw] 
\tikzstyle{medge}=[draw,decoration={markings,mark=at position 0.5 with {\draw(.15, .15) -- (-.15, -.15);\draw(-.15, .15) -- (.15, -.15);}},postaction={decorate}] 
\node (s) at (0, 0) {};
\node (b) at (90:1.5) {};
\node (a) at (-30:1.5) {};
\node (c) at (210:1.5) {};
\path[edge] (a) -- (s);
\path[edge] (a) -- (c);
\path[edge] (s) -- (b);
\path[edge] (s) -- (c);
\path[medge] (a) -- (b);
\path[medge] (b) -- (c);
\end{tikzpicture}$\mathcal{T}_{0,2}$
\begin{tikzpicture}[scale=1]
\useasboundingbox (-1.5,-0.9) rectangle (1.5,1.6);
\tikzstyle{every node}=[draw,circle,inner sep=1.6pt,fill=green!60!black] 
\tikzstyle{edge}=[draw] 
\tikzstyle{medge}=[draw,decoration={markings,mark=at position 0.5 with {\draw(.15, .15) -- (-.15, -.15);\draw(-.15, .15) -- (.15, -.15);}},postaction={decorate}] 
\node (s) at (0, 0) {};
\node (b) at (90:1.5) {};
\node (a) at (-30:1.5) {};
\node (c) at (210:1.5) {};
\path[edge] (a) -- (s);
\path[medge] (a) -- (c);
\path[edge] (s) -- (b);
\path[edge] (s) -- (c);
\path[medge] (a) -- (b);
\path[medge] (b) -- (c);
\end{tikzpicture}$\mathcal{T}_{0,3}$

\vspace{0.5\baselineskip}
\begin{tikzpicture}[scale=1]
\useasboundingbox (-1.5,-1.3) rectangle (1.5,1.6);
\tikzstyle{every node}=[draw,circle,inner sep=1.6pt,fill=green!60!black] 
\tikzstyle{edge}=[draw] 
\tikzstyle{medge}=[draw,decoration={markings,mark=at position 0.5 with {\draw(.15, .15) -- (-.15, -.15);\draw(-.15, .15) -- (.15, -.15);}},postaction={decorate}] 
\node (s) at (0, 0) {};
\node (b) at (90:1.5) {};
\node (a) at (-30:1.5) {};
\node (c) at (210:1.5) {};
\path[name path=p1] (a) -- (c);
\path[name path=p2] (s) .. controls ($ (s) + (-30:0.5) $) and ($ (a) + (230:1.5) $) .. (a);
\path [name intersections={of=p1 and p2,by=x}];
\path[draw] (s) .. controls ($ (s) + (-30:0.5) $) and ($ (a) + (230:1.5) $) .. (a);
\node[circle,fill=white,draw=none,inner sep=2pt] at (x) {};
\path[edge] (a) -- (c);
\path[edge] (s) -- (b);
\path[edge] (s) -- (c);
\path[edge] (a) -- (b);
\path[edge] (b) -- (c);
\end{tikzpicture}$\mathcal{T}_{1,0}$
\begin{tikzpicture}[scale=1]
\useasboundingbox (-1.5,-1.3) rectangle (1.5,1.6);
\tikzstyle{every node}=[draw,circle,inner sep=1.6pt,fill=green!60!black] 
\tikzstyle{edge}=[draw] 
\tikzstyle{medge}=[draw,decoration={markings,mark=at position 0.5 with {\draw(.15, .15) -- (-.15, -.15);\draw(-.15, .15) -- (.15, -.15);}},postaction={decorate}] 
\node (s) at (0, 0) {};
\node (b) at (90:1.5) {};
\node (a) at (-30:1.5) {};
\node (c) at (210:1.5) {};
\path[name path=p1] (a) -- (c);
\path[name path=p2] (s) .. controls ($ (s) + (-30:0.5) $) and ($ (a) + (230:1.5) $) .. (a);
\path [name intersections={of=p1 and p2,by=x}];
\path[draw] (s) .. controls ($ (s) + (-30:0.5) $) and ($ (a) + (230:1.5) $) .. (a);
\node[circle,fill=white,draw=none,inner sep=2pt] at (x) {};
\path[edge] (a) -- (c);
\path[edge] (s) -- (b);
\path[edge] (s) -- (c);
\path[edge] (a) -- (b);
\path[medge] (b) -- (c);
\end{tikzpicture}$\mathcal{T}_{1,1}$
\begin{tikzpicture}[scale=1]
\useasboundingbox (-1.5,-1.3) rectangle (1.5,1.6);
\tikzstyle{every node}=[draw,circle,inner sep=1.6pt,fill=green!60!black] 
\tikzstyle{edge}=[draw] 
\tikzstyle{medge}=[draw,decoration={markings,mark=at position 0.5 with {\draw(.15, .15) -- (-.15, -.15);\draw(-.15, .15) -- (.15, -.15);}},postaction={decorate}] 
\node (s) at (0, 0) {};
\node (b) at (90:1.5) {};
\node (a) at (-30:1.5) {};
\node (c) at (210:1.5) {};
\path[name path=p1] (a) -- (c);
\path[name path=p2] (s) .. controls ($ (s) + (-30:0.5) $) and ($ (a) + (230:1.5) $) .. (a);
\path [name intersections={of=p1 and p2,by=x}];
\path[draw] (s) .. controls ($ (s) + (-30:0.5) $) and ($ (a) + (230:1.5) $) .. (a);
\node[circle,fill=white,draw=none,inner sep=2pt] at (x) {};
\path[edge] (a) -- (c);
\path[edge] (s) -- (b);
\path[edge] (s) -- (c);
\path[medge] (a) -- (b);
\path[medge] (b) -- (c);
\end{tikzpicture}$\mathcal{T}_{1,2}$
\begin{tikzpicture}[scale=1]
\useasboundingbox (-1.5,-1.3) rectangle (1.5,1.6);
\tikzstyle{every node}=[draw,circle,inner sep=1.6pt,fill=green!60!black] 
\tikzstyle{edge}=[draw] 
\tikzstyle{medge}=[draw,decoration={markings,mark=at position 0.5 with {\draw(.15, .15) -- (-.15, -.15);\draw(-.15, .15) -- (.15, -.15);}},postaction={decorate}] 
\node (s) at (0, 0) {};
\node (b) at (90:1.5) {};
\node (a) at (-30:1.5) {};
\node (c) at (210:1.5) {};
\path[name path=p1] (a) -- (c);
\path[name path=p2] (s) .. controls ($ (s) + (-30:0.5) $) and ($ (a) + (230:1.5) $) .. (a);
\path [name intersections={of=p1 and p2,by=x}];
\path[draw] (s) .. controls ($ (s) + (-30:0.5) $) and ($ (a) + (230:1.5) $) .. (a);
\node[circle,fill=white,draw=none,inner sep=2pt] at (x) {};
\path[medge] (a) -- (c);
\path[edge] (s) -- (b);
\path[edge] (s) -- (c);
\path[medge] (a) -- (b);
\path[medge] (b) -- (c);
\end{tikzpicture}$\mathcal{T}_{1,3}$

\vspace{0.5\baselineskip}
\begin{tikzpicture}[scale=1]
\useasboundingbox (-1.5,-1.3) rectangle (1.5,1.6);
\tikzstyle{every node}=[draw,circle,inner sep=1.6pt,fill=green!60!black] 
\tikzstyle{edge}=[draw] 
\tikzstyle{medge}=[draw,decoration={markings,mark=at position 0.5 with {\draw(.15, .15) -- (-.15, -.15);\draw(-.15, .15) -- (.15, -.15);}},postaction={decorate}] 
\node (s) at (0, 0) {};
\node (b) at (90:1.5) {};
\node (a) at (-30:1.5) {};
\node (c) at (210:1.5) {};
\path[name path=p1] (a) -- (c);
\path[name path=p2] (s) .. controls ($ (s) + (-30:0.5) $) and ($ (a) + (230:1.5) $) .. (a);
\path [name intersections={of=p1 and p2,by=x}];
\path[draw] (s) .. controls ($ (s) + (-30:0.5) $) and ($ (a) + (230:1.5) $) .. (a);
\node[circle,fill=white,draw=none,inner sep=2pt] at (x) {};
\path[edge] (a) -- (c);
\path[name path=pa1] (a) -- (b);
\path[name path=pa2] (s) .. controls ($ (s) + (90:0.5) $) and ($ (b) + (-10:1.5) $) .. (b);
\path [name intersections={of=pa1 and pa2,by=xa}];
\path[draw] (s) .. controls ($ (s) + (90:0.5) $) and ($ (b) + (-10:1.5) $) .. (b);
\node[circle,fill=white,draw=none,inner sep=2pt] at (xa) {};
\path[draw] (a) -- (b);
\path[edge] (s) -- (c);
\path[edge] (b) -- (c);
\end{tikzpicture}$\mathcal{T}_{2,0}$
\begin{tikzpicture}[scale=1]
\useasboundingbox (-1.5,-1.3) rectangle (1.5,1.6);
\tikzstyle{every node}=[draw,circle,inner sep=1.6pt,fill=green!60!black] 
\tikzstyle{edge}=[draw] 
\tikzstyle{medge}=[draw,decoration={markings,mark=at position 0.5 with {\draw(.15, .15) -- (-.15, -.15);\draw(-.15, .15) -- (.15, -.15);}},postaction={decorate}] 
\node (s) at (0, 0) {};
\node (b) at (90:1.5) {};
\node (a) at (-30:1.5) {};
\node (c) at (210:1.5) {};
\path[name path=p1] (a) -- (c);
\path[name path=p2] (s) .. controls ($ (s) + (-30:0.5) $) and ($ (a) + (230:1.5) $) .. (a);
\path [name intersections={of=p1 and p2,by=x}];
\path[draw] (s) .. controls ($ (s) + (-30:0.5) $) and ($ (a) + (230:1.5) $) .. (a);
\node[circle,fill=white,draw=none,inner sep=2pt] at (x) {};
\path[edge] (a) -- (c);
\path[name path=pa1] (a) -- (b);
\path[name path=pa2] (s) .. controls ($ (s) + (90:0.5) $) and ($ (b) + (-10:1.5) $) .. (b);
\path [name intersections={of=pa1 and pa2,by=xa}];
\path[draw] (s) .. controls ($ (s) + (90:0.5) $) and ($ (b) + (-10:1.5) $) .. (b);
\node[circle,fill=white,draw=none,inner sep=2pt] at (xa) {};
\path[draw] (a) -- (b);
\path[edge] (s) -- (c);
\path[medge] (b) -- (c);
\end{tikzpicture}$\mathcal{T}_{2,1}$
\begin{tikzpicture}[scale=1]
\useasboundingbox (-1.5,-1.3) rectangle (1.5,1.6);
\tikzstyle{every node}=[draw,circle,inner sep=1.6pt,fill=green!60!black] 
\tikzstyle{edge}=[draw] 
\tikzstyle{medge}=[draw,decoration={markings,mark=at position 0.5 with {\draw(.15, .15) -- (-.15, -.15);\draw(-.15, .15) -- (.15, -.15);}},postaction={decorate}] 
\node (s) at (0, 0) {};
\node (b) at (90:1.5) {};
\node (a) at (-30:1.5) {};
\node (c) at (210:1.5) {};
\path[name path=p1] (a) -- (c);
\path[name path=p2] (s) .. controls ($ (s) + (-30:0.5) $) and ($ (a) + (230:1.5) $) .. (a);
\path [name intersections={of=p1 and p2,by=x}];
\path[draw] (s) .. controls ($ (s) + (-30:0.5) $) and ($ (a) + (230:1.5) $) .. (a);
\node[circle,fill=white,draw=none,inner sep=2pt] at (x) {};
\path[edge] (a) -- (c);
\path[name path=pa1] (a) -- (b);
\path[name path=pa2] (s) .. controls ($ (s) + (90:0.5) $) and ($ (b) + (-10:1.5) $) .. (b);
\path [name intersections={of=pa1 and pa2,by=xa}];
\path[draw] (s) .. controls ($ (s) + (90:0.5) $) and ($ (b) + (-10:1.5) $) .. (b);
\node[circle,fill=white,draw=none,inner sep=2pt] at (xa) {};
\path[medge] (a) -- (b);
\path[edge] (s) -- (c);
\path[medge] (b) -- (c);
\end{tikzpicture}$\mathcal{T}_{2,2}$
\begin{tikzpicture}[scale=1]
\useasboundingbox (-1.5,-1.3) rectangle (1.5,1.6);
\tikzstyle{every node}=[draw,circle,inner sep=1.6pt,fill=green!60!black] 
\tikzstyle{edge}=[draw] 
\tikzstyle{medge}=[draw,decoration={markings,mark=at position 0.5 with {\draw(.15, .15) -- (-.15, -.15);\draw(-.15, .15) -- (.15, -.15);}},postaction={decorate}] 
\node (s) at (0, 0) {};
\node (b) at (90:1.5) {};
\node (a) at (-30:1.5) {};
\node (c) at (210:1.5) {};
\path[name path=p1] (a) -- (c);
\path[name path=p2] (s) .. controls ($ (s) + (-30:0.5) $) and ($ (a) + (230:1.5) $) .. (a);
\path [name intersections={of=p1 and p2,by=x}];
\path[draw] (s) .. controls ($ (s) + (-30:0.5) $) and ($ (a) + (230:1.5) $) .. (a);
\node[circle,fill=white,draw=none,inner sep=2pt] at (x) {};
\path[medge] (a) -- (c);
\path[name path=pa1] (a) -- (b);
\path[name path=pa2] (s) .. controls ($ (s) + (90:0.5) $) and ($ (b) + (-10:1.5) $) .. (b);
\path [name intersections={of=pa1 and pa2,by=xa}];
\path[draw] (s) .. controls ($ (s) + (90:0.5) $) and ($ (b) + (-10:1.5) $) .. (b);
\node[circle,fill=white,draw=none,inner sep=2pt] at (xa) {};
\path[medge] (a) -- (b);
\path[edge] (s) -- (c);
\path[medge] (b) -- (c);
\end{tikzpicture}$\mathcal{T}_{2,3}$
\caption{Maps derived from the embeddings of the graph $K_4$, with and without twists. Only $9$ of the $12$ drawings shown here are distinct, 
as $(\mathcal{T}_{0,3}, \mathcal{T}_{2,3})$, $(\mathcal{T}_{1,1}, \mathcal{T}_{2,2})$, and $(\mathcal{T}_{1,2}, \mathcal{T}_{2,1})$ are isomorphic pairs.
Conventions for edges as in Figure~\ref{fig:blue}.
}
\end{figure}
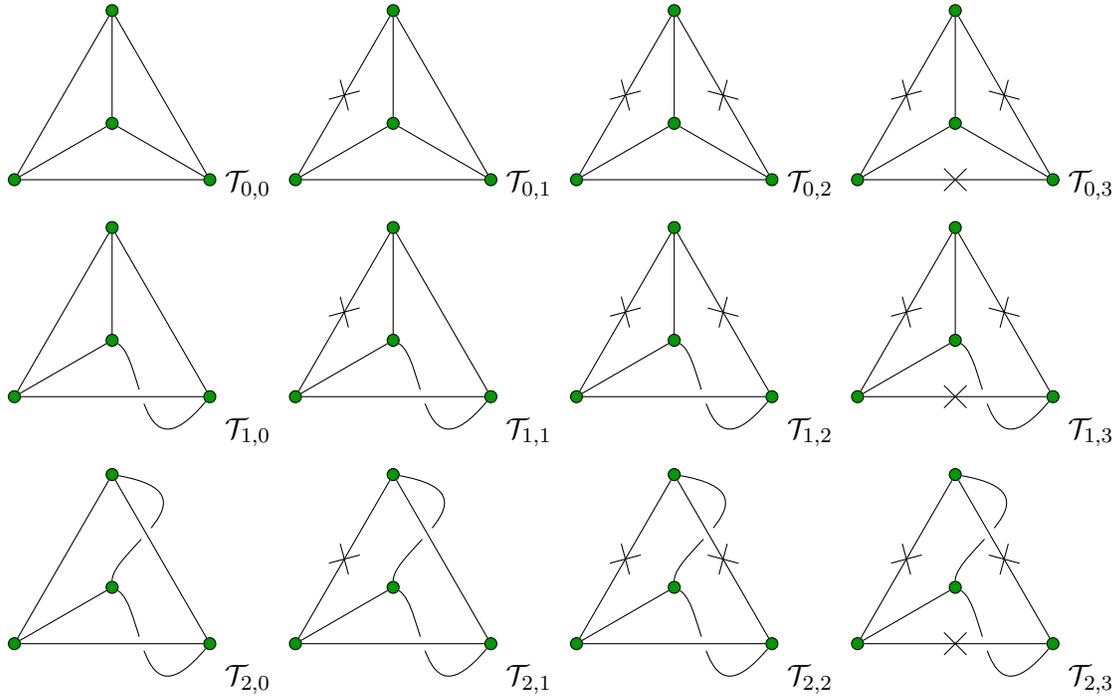

\begin{enumerate}[label=(\roman*)]
\item For the theta graph for all distinct embeddings and sets of twists:
\begin{align}
K(\mathcal{M}_{6,0}^{2}; l) & = {\left(l + 1\right)}^{3} + 8 \\
K(\mathcal{M}_{6,1}^{2}; l) & = {\left(l + 1\right)}^{3} + 4 \\
K(\mathcal{M}_{7,0}^{2}; l) & = {\left(l + 1\right)}^{3} + 3 \, (l + 1) + 2 \\
K(\mathcal{M}_{7,1}^{2}; l) & = {\left(l + 1\right)}^{3} + (l + 1) + 2
\end{align}

\item For the tetrahedron for all distinct embeddings and sets of twists:
\begin{align}
K(\mathcal{T}_{0,0}; l) & = {\left(l + 1\right)}^{6} + 4 \, {\left(l + 1\right)}^{3} + 3 \, {\left(l + 1\right)}^{2} \\
K(\mathcal{T}_{0,1}; l) & = {\left(l + 1\right)}^{6} + 4 \, {\left(l + 1\right)}^{3} + 3 \, {\left(l + 1\right)}^{2} + 2 \, (l + 1) \\
K(\mathcal{T}_{0,2}; l) & = {\left(l + 1\right)}^{6} + 4 \, {\left(l + 1\right)}^{3} + 2 \, {\left(l + 1\right)}^{2} + 3 \, (l + 1) \\
K(\mathcal{T}_{0,3}; l) = K(\mathcal{T}_{2,3}; l) & = {\left(l + 1\right)}^{6} + 5 \, {\left(l + 1\right)}^{3} + 3 \, {\left(l + 1\right)}^{2} + 6 \, (l + 1) + 2 \\
K(\mathcal{T}_{1,0}; l) & = {\left(l + 1\right)}^{6} + 4 \, {\left(l + 1\right)}^{3} + 3 \, (l + 1) \\
K(\mathcal{T}_{1,1}; l) = K(\mathcal{T}_{2,2}; l) & = {\left(l + 1\right)}^{6} + 5 \, {\left(l + 1\right)}^{3} + {\left(l + 1\right)}^{2} + 2 \, (l + 1) + 2 \\
K(\mathcal{T}_{1,2}; l) = K(\mathcal{T}_{2,1}; l) & = {\left(l + 1\right)}^{6} + 4 \, {\left(l + 1\right)}^{3} + 2 \, {\left(l + 1\right)}^{2} + 2 \, (l + 1) + 2 \\
K(\mathcal{T}_{1,3}; l) & = {\left(l + 1\right)}^{6} + 4 \, {\left(l + 1\right)}^{3} + 5 \, {\left(l + 1\right)}^{2} + 4 \, (l + 1) + 4 \\
K(\mathcal{T}_{2,0}; l) & = {\left(l + 1\right)}^{6} + 4 \, {\left(l + 1\right)}^{3} + 4   
\end{align}
\item For the cube in the usual embedding, $\Sigma$  (see Figure~\ref{fig:threeembeddings}(a)), on the sphere, with no twist and one twisted edge, 
and for the untwisted toroidal embeddings, $\Sigma'$ and $\Sigma''$ (see Figures~\ref{fig:threeembeddings}(b) and \ref{fig:threeembeddings}(c)), respectively:
\begin{align}
K(\Sigma_0; l) & = {\left(l + 1\right)}^{12} + 8 \, {\left(l + 1\right)}^{9} + 32 \, {\left(l + 1\right)}^{6} + 64 \, {\left(l + 1\right)}^{3} + 64 \\
K(\Sigma_1; l) & = {\left(l + 1\right)}^{12} + 8 \, {\left(l + 1\right)}^{9} + 26 \, {\left(l + 1\right)}^{6} + 40 \, {\left(l + 1\right)}^{3} + 32 \\
K(\Sigma'_0; l) & = {\left(l + 1\right)}^{12} + 8 \, {\left(l + 1\right)}^{9} + 2 \, {\left(l + 1\right)}^{8} + 24 \, {\left(l + 1\right)}^{6} + 8 \, {\left(l + 1\right)}^{5}  \\
& \quad {} + 17 \, {\left(l + 1\right)}^{4} + 32 \, {\left(l + 1\right)}^{3} + 8 \, {\left(l + 1\right)}^{2} + 16 \notag \\
K(\Sigma''_0; l) & = {\left(l + 1\right)}^{12} + 8 \, {\left(l + 1\right)}^{9} + 6 \, {\left(l + 1\right)}^{8} + 16 \, {\left(l + 1\right)}^{6} + 24 \, {\left(l + 1\right)}^{5}  \\
& \quad {} + 9 \, {\left(l + 1\right)}^{4} + 16 \, {\left(l + 1\right)}^{3} + 24 \, {\left(l + 1\right)}^{2} + 16 \notag
\end{align}

\item For the dodecahedron on the sphere, with no twist and one twisted edge, respectively:
\begin{align}
K(\mathcal{D}_0; l) & = {\left(l + 1\right)}^{30} + 20 \, {\left(l + 1\right)}^{27} + 160 \, {\left(l + 1\right)}^{24} + 660 \, {\left(l + 1\right)}^{21} + 36 \, {\left(l + 1\right)}^{20} \\
 & \quad {} + 1510 \, {\left(l + 1\right)}^{18} + 360 \, {\left(l + 1\right)}^{17} + 1972 \, {\left(l + 1\right)}^{15} + 1260 \, {\left(l + 1\right)}^{14} \notag \\
 & \quad {} + 120 \, {\left(l + 1\right)}^{13} + 1560 \, {\left(l + 1\right)}^{12} + 1800 \, {\left(l + 1\right)}^{11} + 636 \, {\left(l + 1\right)}^{10} \notag \\
 & \quad {} + 660 \, {\left(l + 1\right)}^{9} + 1020 \, {\left(l + 1\right)}^{8} + 600 \, {\left(l + 1\right)}^{7} + 125 \, {\left(l + 1\right)}^{6} \notag \\
K(\mathcal{D}_1; l) & = {\left(l + 1\right)}^{30} + 20 \, {\left(l + 1\right)}^{27} + 160 \, {\left(l + 1\right)}^{24} + 2 \, {\left(l + 1\right)}^{23} + 2 \, {\left(l + 1\right)}^{22} \\
 & \quad {} + 660 \, {\left(l + 1\right)}^{21} + 52 \, {\left(l + 1\right)}^{20} + 24 \, {\left(l + 1\right)}^{19} + 1512 \, {\left(l + 1\right)}^{18}  \notag \\
 & \quad{} + 394 \, {\left(l + 1\right)}^{17} + 124 \, {\left(l + 1\right)}^{16} + 1984 \, {\left(l + 1\right)}^{15} + 1250 \, {\left(l + 1\right)}^{14}  \notag \\
 & \quad{} + 428 \, {\left(l + 1\right)}^{13} + 1608 \, {\left(l + 1\right)}^{12} + 1738 \, {\left(l + 1\right)}^{11} + 936 \, {\left(l + 1\right)}^{10}  \notag \\
 & \quad{} + 848 \, {\left(l + 1\right)}^{9} + 984 \, {\left(l + 1\right)}^{8} + 708 \, {\left(l + 1\right)}^{7} + 285 \, {\left(l + 1\right)}^{6} + 50 \, {\left(l + 1\right)}^{5}  \notag
\end{align}
We note in passing that the corresponding formula for the untwisted embedding of the Petersen graph in the projective plane, which is admittedly of less chemical interest, is
\begin{equation}
K(\text{Petersen} ;l) = {\left(l + 1\right)}^{15} + 10 \, {\left(l + 1\right)}^{12} + 30 \, {\left(l + 1\right)}^{9} + 55 \, {\left(l + 1\right)}^{6} + 55 \, {\left(l + 1\right)}^{3}
\end{equation}

\item For the $k$-prism $\mathcal{R}^k$:

For $k$ odd, the prism with linear polyacene motifs, embedded on the sphere without a twist has
\begin{equation}
K(\mathcal{R}^k_0;l) =  ((l+1)\sp3 + 2)\sp{k} - 2\sp{k}
\end{equation}
and for $k$ even it has
\begin{equation}
K(\mathcal{R}^k_0; l) = [ ((l+1)\sp3 + 2)\sp{k/2} + 2\sp{k/2} ]\sp2 %. 
\end{equation}
\end{enumerate}

We note that as the rules (a) to (c) apply without change to the limiting case of $l = 0$ hexagons in the linear polyene chain, 
the formulas
for the untwisted chains apply to the leapfrog \cite{PWFTP1994} of an embedded cubic graph and hence give the Kekul\'{e} count 
of the leapfrog by summation of coefficients of all powers of $(l+1)$. 
For example, the formula for the untwisted dodecahedron $\mathcal{D}_0$ gives the number of Kekul\'{e} structures of the icosahedral C$_{60}$ fullerene as $K(\mathcal{D}_0; 0)  = 12500$.
Simple results are also found for the numbers of perfect matchings of leapfrog prisms, namely $3\sp{k}-2\sp{k}$ and $(3\sp{k/2}+2\sp{k/2})\sp2$.
The prism itself has Kekul{\'e} count given by sequence A068397 \cite{oeisA068397}, i.e.\ $K$ for the $k$-prism is $F(k+1) + F(k-1) + (-1)^k + 1$. 

\section{Structural calculations}

\begin{figure}[!htb]
\centering
\includegraphics[scale=0.8]{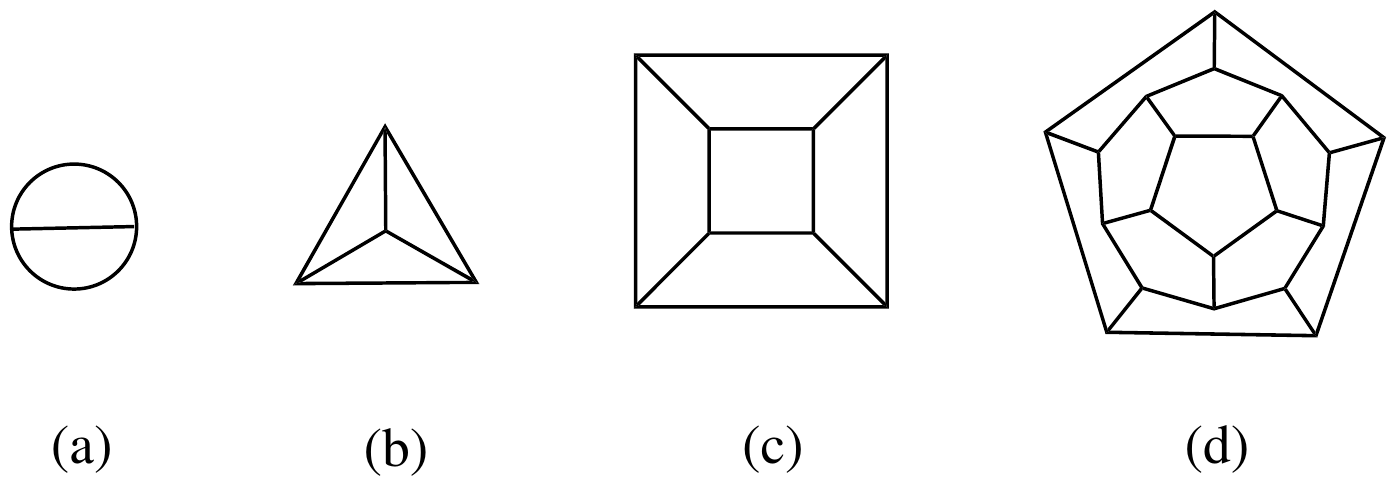}
\caption{Cubic graphs used as the basis for flat hexagonal complexes: (a) the theta graph, (b) the tetrahedron, (c) the cube, (d) the dodecahedron. In cases (b) to (d), each edge is decorated with an anthracene chain; for case (a) see Figures~\ref{fig:PWF2}, \ref{fig:PWF3} and \ref{fig:PWF5}.}
\label{fig:PWF1}
\end{figure}

This section connects the foregoing mathematical development with possible realisations of new unsturated hydrocarbon frameworks.
Graph theoretical considerations based on Kekulé counts, HOMO-LUMO gap and $\pi$ energy can be valuable indicators of stability of a $\pi$ system.  
In particular, it is useful to know if a $\pi$ system is predicted to have
a closed shell (in which all electrons are paired) and to
characterise such shells in terms  of whether this electron configuration is  {\it properly closed}
(having all bonding orbitals occupied and all antibonding orbitals empty).
HOMO-LUMO maps \cite{FoPi2010,JaFoPi2012} can give a useful picture of
trends in frontier-orbital energies
and shell types for families of molecules.
However, for a more reliable estimate of the prospects for overall stability of an unsaturated hydrocarbon C$_x$H$_y$ it is 
necessary to take into account the full range of steric and electronic effects arising from 
both $\sigma$ and $\pi$ electronic subsystems. 
This section reports a selection of
preliminary all-electron structural calculations using standard quantum chemical methods for various examples of chemical 
hexagonal complexes. They serve to show that this generalisation of benzenoids is chemically as well as mathematically 
plausible, and give clues to some of the factors that can influence absolute and relative stabilities. 
The systems chosen 
for study are linear polyacene expansions of the three cubic Platonic polyhedra, and a wider choice of isomeric 
expansions of the simplest cubic graph, the theta graph (see Figure~\ref{fig:PWF1}).
All these structures correspond to 
the standard  embedding on the sphere; twisted systems and alternative embeddings were 
left for future investigations.

In each case, the structure was optimised at the DFT level using the B3LYP functional and, in all but one, 
the 6-31G* basis for C and 6-31G for H. (In the case of the dodecahedral complex, C$_{420}$H$_{180}$, 
the basis was reduced to 6-31G for all atoms, on grounds of computational cost.) Candidate minima were 
checked in most cases in the usual way, by diagonalization of the Hessian. (For the largest cases, of the 
expanded cube and dodecahedron, stability of the candidate minimum structure was checked by relaxation of several
nearby unsymmetrically perturbed structures.) Calculations were carried out with the QChem and Gaussian~16 
packages \cite{qchem4,g16}. Energies and lowest harmonic frequencies are 
reported in Tables~\ref{tab:PWF1} and \ref{tab:PWF2}, together with geometric 
parameters and three graph invariants (Kekul{\'e} count, H{\"u}ckel binding energy per carbon
atom, and H{\"u}ckel HOMO-LUMO gap). Snapshots of some optimised structures are shown in Figure~\ref{fig:PWF6}.

\begin{table}[!t]
\centering
\setlength{\tabcolsep}{3pt}
\begin{tabular}{lclrcrrc}
Base graph & $l$ & Formula & $K$ & $(E_\pi/n)\beta$ & $\Delta_{HL}\beta$& $E$/eV & $\tilde{\nu}/\text{cm}^{-1}$ \\
\hline 
Theta graph    & $3$ & C$_{ 42}$H$_{ 18}$   & 72 & $1.439933$ & $0.54778$ & $-43836.415$ & $76$  \\
                      & $4$ & C$_{ 54}$H$_{ 24}$  & 133 & $1.431856$ & $0.39425$ & $-56381.449$ & $52$  \\
                      & $5$ & C$_{ 66}$H$_{ 30}$  & 224 & $1.426646$ & $0.29932$ &$-68925.208$ & $34$  \\
Tetrahedron     & $3$ & C$_{ 84}$H$_{ 36}$  & 4356 & $1.439017$ & $0.56885$ & $-175391.522$ & $39$ \\
Cube               & $3$ & C$_{168}$H$_{ 72}$  & 19009600 & $1.443904$ & $0.54778$ & $-175390.812$ & ---  \\
Dodecahedron  & $3$ & C$_{420}$H$_{180}$ & 1561300213688815616 % $2^{12} \cdot 109 \cdot 281^3 \cdot 397^2$
& $1.439049$ & $0.55627$ &$-438398.859$ & ---   
\end{tabular}
\caption{Hexagonal chemical complexes based on inflation
of cubic graphs with linear polyacene chains along edges. All isomers are based on linear annelation to alternate edges of the branching hexagons corresponding to vertices of the cubic graph. 
$l$ is the length of the chain motif,
$K$ is the number of Kekul{\'e} structures, $(E_\pi/n)$ is the H{\"u}ckel $\pi$ energy per carbon centre, 
in units of the  $\beta$ resonance parameter,  $\Delta_{HL}$ is the H{\"u}ckel HOMO-LUMO gap, in the same units, $E$ is the total all-electron energy in eV (see text for the level of theory), 
and $\tilde{\nu}$ is the wavenumber in units of cm$^{-1}$ of the vibrational mode of lowest energy.}
\label{tab:PWF1}
\end{table}

\begin{table}[!t]
\centering
\begin{tabular}{llrcclrr}
 Formula & Isomer & $K$ & $(E_\pi/n)\beta$ & $\Delta_{HL}\beta$ & $E$/eV & $\Delta E$/eV & $\tilde{\nu}/\text{cm}^{-1}$   
  \\
\hline 
 C$_{42}$H$_{18}$ 
    & Aa & 72    & $1.439933$ & $0.54778$ & $-43836.415$ & 1.789 & $76$   \\
    & Ab & 108  & $1.445495$ & $0.81078$ & $-43837.062$ & 1.142 & $117$  \\
    & Ac & 144  & $1.450217$ & $0.85153$ & $-43836.830$ & 1.373 & $ 70$ \\
    & Ad & 144  & $1.450269$ & $0.85153$ & $-43836.782$ & 1.422 & $ 118$ \\
    & Pa & 208  & $1.455866$ & $1.09287$ & $-43834.363$ &  3.840& $ 74$ \\ 
    & Pb & 160  & $1.450380$ & $1.09287$ & $-43838.204$ & 0.000  & $ 117$ \\
    & Pc & 208  & $1.456138$ & $1.09287$ & $-43836.842$ &  1.362& $ 124$ \\
    & Pd & 176  & $1.452603$ & $1.09287$ & $-43836.495$ &  1.708& $ 102$ \\
    & Pe & 208  & $1.455940$ & $1.09287$ & $-43835.822$ &  2.382& $ 136$ \\
    & Pf & 176  &  $1.452827$ & $1.03801$ & $-43837.000$ &  1.204& $ 102$ \\
  C$_{66}$H$_{30}$ 
   & Fa & 224   & $1.426646$ & $0.29932$ & $-68925.208$ &  2.610& $34$ \\
   & Fb & 1088 & $1.436307$ & $0.75569$ & $-68927.219$ &  0.599 & $52$ \\
   & Fc & 2500 & $1.444639$ & $0.91215$ & $-68927.818$ &  0.000& $50$ 
\end{tabular}
\caption{Attachment isomers of hexagonal chemical complexes based on inflation
of the theta graph with catafusenes composed of $3$ and $5$ hexagons
(for formulas C$_{42}$H$_{18}$ and C$_{66}$H$_{30}$, respectively). 
The isomers are depicted in Figures~\ref{fig:PWF3} and \ref{fig:PWF5}.
$K$ is the number of Kekul{\'e} structures, $(E_\pi/n)$ is the H{\"u}ckel $\pi$ energy per carbon centre, 
in units of the  $\beta$ resonance parameter,  $\Delta_{HL}$ is the H{\"u}ckel HOMO-LUMO gap, in the same units, $E$ is the total all-electron energy in eV (see text for the level of theory), 
and $\tilde{\nu}$ is the wavenumber in units of cm$^{-1}$ of the vibrational mode of lowest energy.
}
\label{tab:PWF2}
\end{table}

Structures based on the theta graph (Figure \ref{fig:PWF1}(a)) with all edges inflated to linear polyacene 
chains of $l$ hexagons were optimised successively for chains of length $l = 3, 4, 5$. These correspond to 
molecular formulas C$_{42}$H$_{18}$, C$_{54}$H$_{24}$, and C$_{66}$H$_{30}$, respectively. 
All were found to occupy minima on the potential energy surface within the model chemistry. 
The optimised structures are barrels, with CC bond lengths in the expected range for polycyclic 
aromatic systems in this model chemistry, 
\begin{figure}[!p]
\centering
\includegraphics[scale=0.7]{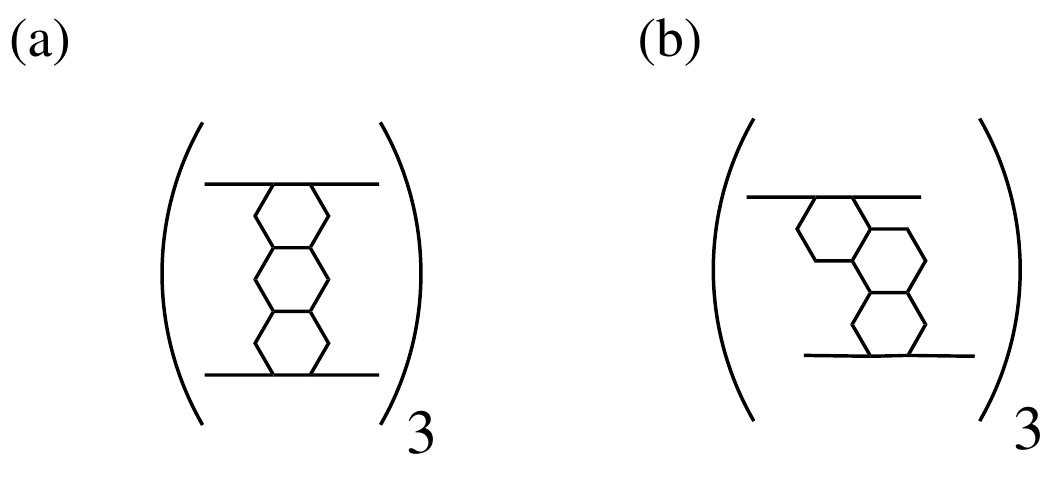}
\caption{A {\lq polymer\rq} molecular notation for (a) anthracene and (b) phenanthrene expansions of the theta graph into flat hexagonal complexes of formula C$_{42}$H$_{18}$. Each monomer is to be repeated twice more to give a cyclic molecular structure with three-fold symmetry, topped and tailed by a hexagonal ring. Only graph vertices corresponding to carbon atoms are shown: vertices of degree two each carry a single H atom, and the graph is filled out with an appropriate Kekulé system of double bonds. 
The illustrated isomers are those denoted Aa and Pf, respectively, in the notation of Figure~\ref{fig:PWF3}.}
\label{fig:PWF2}
\end{figure}
\begin{figure}[!p]
\centering
\includegraphics[scale=0.7]{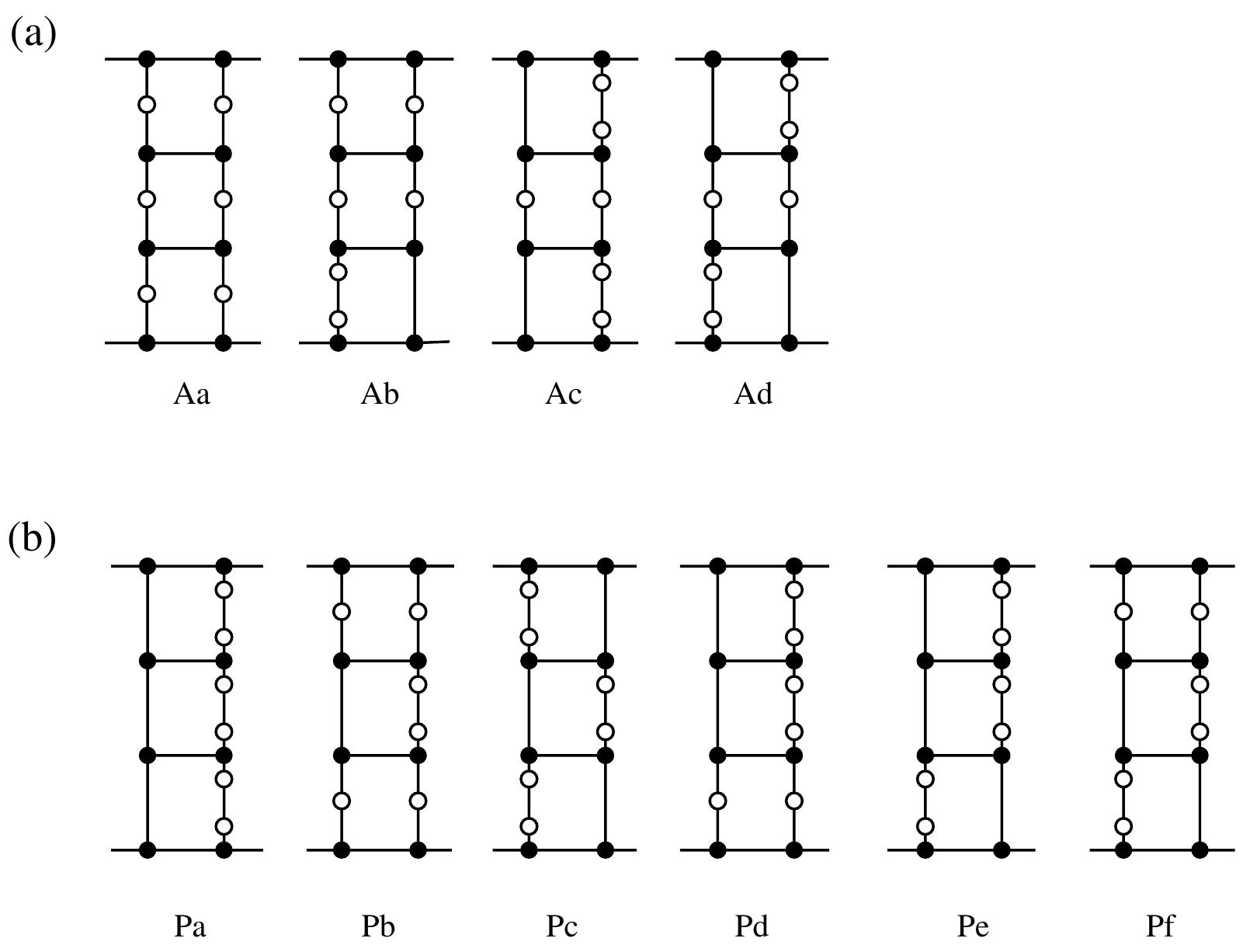}
\caption{Schematic notation for attachment isomers of three-hexagon catafusene expansions of the theta graph. Each vertical block represents a possible strip in a three-fold symmetric isomer based on either anthracene (A) or phenanthrene (P). Notation: a black circle denotes a C centre that has three C neighbours; a white circle denotes a C centre that has two C neighbours and one H neighbour.  For simplicity, the catafusene strip is shown as vertical; in the molecule the strip must bend in order to keep parallel the median planes of the hexagons centred on the C$_3$ axis of the hexagonal complex. Double bonds can be added, for example in any way consistent with internal Kekulé structures of the parent catafusene.}
\label{fig:PWF3}
\end{figure}
and low-frequency vibrational modes consistent with the flexibility expected of their open cage structures. 
Attempts to optimise the putative molecule with $l = 2$, C$_{30}$H$_{12}$,
failed to yield converged structures that corresponded to the initial molecular graph.
The equivalent inflations using $l = 3$ with the three cubic polyhedra  (Figure \ref{fig:PWF1})
were also used to generate optimised molecular structures with 
formulas C$_{84}$H$_{36}$, C$_{168}$H$_{72}$, and C$_{420}$H$_{180}$, respectively 
(see Table \ref{tab:PWF1} and Figure \ref{fig:PWF1}).

\begin{figure}[!b]
\centering
\includegraphics[scale=0.7]{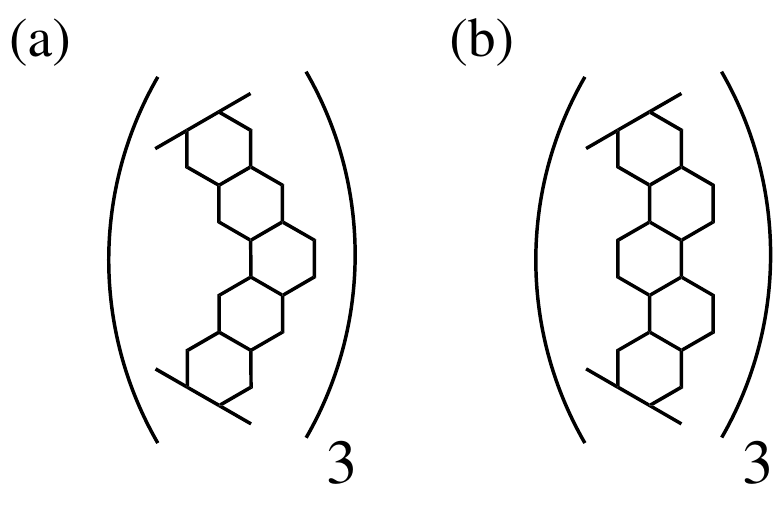}
\caption{Two isomers of hexagonal complexes based on decoration of the theta graph with five-hexagon catafusenes. The diagrams represent three-fold symmetric decorations of the graph with (a) a singly kinked polyacene chain (Fb), and (b) a zig-zag fibonacene chain (Fc). Double bonds can be filled in {\it ad lib} to correspond with internal perfect matchings of the respective catafusenes. Both isomers share molecular formula C$_{66}$H$_{30}$ with the straight-chain isomer~(Fa).}
\label{fig:PWF5}
\end{figure}

Whilst these results are already encouraging, it is not to be expected that the use of the mathematically 
simple linear polyacene fragments to inflate graph edges will automatically lead to the isomer 
of the hexagonal complex that has the highest chemical stability. There are at least two 
variations to the construction recipe that might be expected on electronic and/or steric 
grounds to improve stability. The more obvious of these is that for $l \ge 3$ we have 
choice for the isomer of the catafused benzenoid to be used in the inflation procedure.  
Considered as isolated molecules, bent catafusenes are typically more stable than their 
linear counterparts. Figure \ref{fig:PWF2} illustrates this degree of freedom for the $l = 3$ 
expansions of the theta graph, where phenanthrene offers a plausible alternative to anthracene. 
However, even once we fix on a given catafusene as our 
favoured structural motif, there are still different 
possibilities for its mode of 
annelation to the branch-point hexagons (of type $A_3$) that represent the vertices of the 
original cubic graph. We can, for example, construct {\lq attachment isomers\rq} by choosing 
any contiguous pair $-$CH$-$CH$-$ on the catafusene perimeter as the site of the shared 
connection with the branch-point hexagonal ring. Figure \ref{fig:PWF3} illustrates the variety 
of possible attachment isomers for the case of three-hexagon chains, with the added constraint  
of threefold rotational symmetry around the branching hexagons.

Optimisation shows that both variations on the basic recipe for construction are significant 
(see Table~\ref{tab:PWF2}).
Compared to direct linear annelation, the linear anthracene fragment gives a more stable isomer 
when attached to the branching hexagons in non-linear fashion (Ab).  However, a 
further improvement in total energy comes from
switching to the phenanthrene motif in the construction of the complex, again with a significant 
energetic preference for one particular attachment mode (Pb).

\begin{figure}[!b]
\centering
\begin{minipage}[b]{.5\linewidth}
\centering
\includegraphics[scale=1]{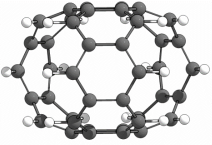}
\subcaption{}\label{fig:1a}
\end{minipage}%
\begin{minipage}[b]{.5\linewidth}
\centering
\includegraphics[scale=0.36]{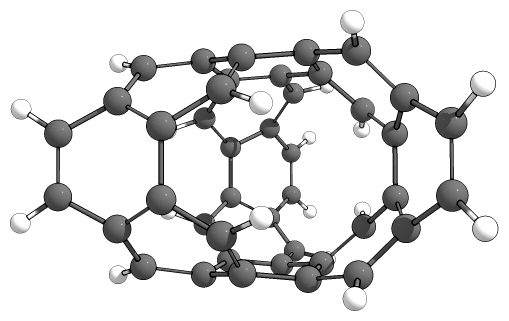}
\subcaption{}\label{fig:1b}
\end{minipage}

\vspace{0.5cm}
\begin{minipage}[b]{.5\linewidth}
\centering
\includegraphics[scale=1]{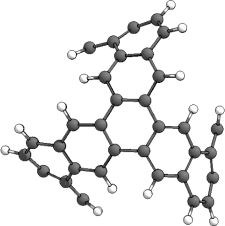}
\vspace{0.7cm}
\subcaption{}\label{fig:1b}
\end{minipage}%
\begin{minipage}[b]{.5\linewidth}
\centering
\includegraphics[scale=1]{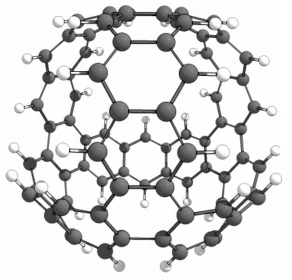}
\subcaption{}\label{fig:1b}
\end{minipage}
\caption{Ball-and-stick representations of some optimised molecular structures based on hexagonal complexes.
(a) Isomer of C$_{42}$H$_{18}$ based on the anthracene expansion of the theta graph shown in Figure \ref{fig:PWF2}(a); 
(b) Isomer of C$_{42}$H$_{18}$ based on the phenanthrene expansion Pb of the theta graph
(see Figure~\ref{fig:PWF3}(b); 
(c) Top view of C$_{66}$H$_{30}$ isomer based on the fibonacene expansion of the theta graph shown in Figure \ref{fig:PWF5}(b);
(d) View down the two-fold axis of a hexagonal complex, C$_{84}$H$_{36}$, based on the anthracene expansion of the
tetrahedron.}
\label{fig:PWF6}
\end{figure}

Preference for a phenanthrene over an anthracene motif is consistent with the relative stabilities of the
isomeric C$_{14}$H$_{6}$ compounds \cite{JChemThermo1979}.
Higher stability of the bent polyacene system is attributed in part to $\pi$
resonance effects, although these are hard to quantify uniquely
\cite{CiesielskiArkadiusz2008WAtK,Cyra}, and are offset by steric effects, such as
H-H repulsion in the bay region.
The computed relative energy of $1.142$~eV ($\sim 110$~kJ~mol$^{-1}$)
of isomers Ab and Pb, each containing three copies of the respective motifs, is compatible with
the differential stability estimated from 
the difference  of $23$~kJ~mol$\sp{-1}$ in the standard formation enthalpies of the pure  
compounds \cite{JChemThermo1979}, but points to a significant role for relief of steric crowding 
in the more open structure of the phenanthrene hexagonal complex Pb 
(see Figure~\ref{fig:PWF3} (b)).
In support of this hypothesis of a major role for steric effects, we note that
the indicators of pure $\pi$ electronic stability do not show any clear correlation
with the computed relative energies of attachment isomers (Table~\ref{tab:PWF2}, Aa-Ad, Pa-Pf). For example, 
whilst it is true that the isomer Pb, which has the lowest all-electron energy, has a higher
Kekul{\'e} count, larger $\pi$ energy per electron and bigger HOMO-LUMO gap than its nearest
competitor, Ab, it does not stand out on any of qualitative measures from the mass of
the phenanthrenoid and anthracenoid isomers. Again, Pb has a large
but not maximum Fries number.  An overall trend to lower
$\pi$ energy per electron and smaller HOMO-LUMO gap, countered by a rapidly increasing
Kekul{\'e} count, is evident for CHCCs with longer
linear polyacene motifs, and frequency calculations suggest increasing
flexibility in larger cages with longer catafusene motifs.

These considerations may also be promising for the prospects of
larger
hexagonal complexes based on cubic polyhedra, where the face sizes are typically larger, and there 
should be more room for avoidance of steric clashes.  With larger faces and longer chains, 
the complexes with twisted M{\"o}bius catafusenes along polyhedral edges 
may also become less sterically disfavoured. 
There are clearly many
possibilities to be explored.
Although by no means complete, this short survey has shown that 
at least some generalised 
hexagonal complexes
survive the initial test of chemical plausibility in that they occupy minima
on the potential surface. Synthetic accessibility is of course another matter.

\section*{Acknowledgements}

TP is supported in part by the Slovenian Research Agency (research program P1-0294 and research projects J1-2481, N1-0032, J1-9187, J1-1690 and N1-0140), 
and in part by H2020 Teaming InnoRenew CoE.

NB is supported in part by the Slovenian Research Agency (research program P1-0294 and research projects J1-2481, J1-9187, J1-1691, and N1-0140.).

CSA is supported by the U.S. Department of Energy, Office of Science, Basic Energy Sciences, under Award \#DE-SC0019394, as part of the Computational Chemical Sciences Program. CSA performed calculations using resources of the National Energy Research Scientific Computing Center (NERSC), a U.S.~Department of Energy Office of Science User Facility operated under Contract No.~DE-AC02-05CH11231.
CSA would also like to thank J. R. R. Verlet for access to computing facilities.

\bibliographystyle{plain}
\bibliography{references}

\end{document}